\documentclass[aps,prc,preprint,superscriptaddress,showpacs,floatfix]{revtex4}
\usepackage{graphicx,color}
\usepackage{braket}
\usepackage{amsmath}
\usepackage{amssymb}
\newlength{\figwidth}

\def\m@thcombine#1#2{%
  \setbox0=\hbox{$#1$}
  \setbox1=\hbox{$#2$}
  \ifdim\wd0>\wd1
    \setbox0=\hbox to\wd1{\hss\box0\hss}
  \else
    \setbox1=\hbox to\wd0{\hss\box1\hss}
  \fi
  \mathop{\vcenter{
    \offinterlineskip\box0\box1}}}
\def\lesim{\m@thcombine<\sim}
\def\gesim{\m@thcombine>\sim}

\begin{document}
\setlength{\figwidth}{0.98\columnwidth}

\title{$s$-wave quasiparticle resonance in neutron-rich drip-line nuclei}

\author{Yoshihiko Kobayashi}
\affiliation{Faculty of Arts and Science, Kyushu University, Fukuoka 819-0395, Japan}
\author{Masayuki Matsuo}
\affiliation{Department of Physics, Faculty of  Science, Niigata University, Niigata 950-2181, Japan}

\begin{abstract}
We investigate unbound single-particle states in pair-correlated drip-line nuclei by describing a low-energy elastic scattering of a neutron in the $s$-wave within the framework of the coordinate space Hartree-Fock-Bogolibov (Bogoliubov-de Genne) equation. Numerical study is performed for a neutron drip-line carbon isotope where the neutron $2s_{1/2}$ orbit is located close to zero energy. Analyzing the S-matrix poles of the elastic scattering, we discuss properties of the $s$-wave quasiparticle resonance and, in particular, behaviors characteristic to drip-line nuclei. It is found that the S-matrix has {\it two pairs of poles}; one pair appears as either a weakly bound state, a virtual state or a resonance while the other pair gives contribution analogous to a bound single-particle state. The $s$-wave quasiparticle resonance emerges with a large variation depending on the pairing gap and the single-particle energy of the $s$-orbit.
\end{abstract}

%\date{\today}
%\pacs{24.50.+g, 25.60.-t, 25.70.-z, 27.40.+z}
\maketitle

\section{Introduction}
The pairing correlation is one of the most important many-body correlations in finite nuclei~\cite{Dean2003, BrinkBroglia, BrogliaZelevinsky}. Condensation of nucleon Cooper pairs causes configuration mixing around the Fermi energy, and affects various aspects of low-lying nuclear structure and low-energy nuclear reaction. The pairing correlation is expected to influence also the properties of exotic nuclei close to the drip-lines, in more characteristic ways than in nuclei close to the stability line. This is because, due to the shallow Fermi energy, the Cooper pairs and the configuration mixing involve weakly-bound and continuum (unbound) single-particle orbits, which have very different features, e.g. spatially extended wave functions. For example the pairing correlation has been discussed extensively as a mechanism to govern the binding and the size of the two-neutron halo nuclei~\cite{Tanihata1985, Hansen1987, Bertsch1991, Dobaczewski1996, Meng1996, Bennaceur2000, Dobaczewski2001, Barranco2001, Myo2002, Tanihata2013, Chen2014, Meng2015, Zhang2017, Hagino2017}.

The single-particle motion of nucleon is also affected by the pairing correlation as is known in the Bogoliubov's quasiparticle theory~\cite{RingSchuck}. Treating properly the asymptotic boundary condition on the quasiparticle wave function far outside the nucleus, one can describe not only bound (and discrete) quasiparticle states but also unbound quasiparticle states with continuum spectra~\cite{Belyaev1987, Bulgac1980,Dobaczewski1984}. In other words, the Bogoliubov's quasiparticle theory allows us to describe a nucleon scattering on a pair-correlated nucleus. As an important consequence, the theory predicts appearance of a novel single-particle resonance, called {\it quasiparticle resonance}~\cite{Belyaev1987,Bulgac1980,Dobaczewski1996,Bennaceur1999}. It arises from a coupling of a bound hole state and unbound continuum states, which is caused by formation of a Cooper pair. As discussed in previous theoretical studies, however, the quasiparticle resonance may be hard to be observed in stable nuclei, where the threshold excitation energy for the unbound single-particle motion (i.e. one-nucleon separation energy) is high and the resonance is likely to be smeared by background complex configurations~\cite{Dobaczewski1996, Bertsch1983}. On the contrary, in nuclei with a small one-nucleon separation energy, in particular, in neutron-rich nuclei near the drip-line, the threshold excitation energy for neutron separation is low, and one can expect better opportunities of observing and exploring the quasiparticle resonance in low-energy continuum spectrum of unbound neutron. The purpose of our studies is, therefore, to reveal theoretically properties of the quasiparticle resonance which emerges in neutron-rich nuclei with focus on the role of the pairing correlation and the weak binding of neutrons.

The quasiparticle resonance, or unbound scattering quasiparticle states in general, is expected to depend on the strength of the pairing correlation, and the position of the Fermi energy as well as the single-particle eigenstates orbiting about the nucleus~\cite{Belyaev1987}. In a preceding work~\cite{Kobayashi2016}, we have studied the $p$-wave quasiparticle resonance of neutron which might emerge in near-neutron-drip-line nuclei around Si region. It is found that the pairing effect on the quasiparticle resonance has a variety depending on whether the $p$-wave single-particle state is located below the Fermi energy (hole-like configuration) or above the Fermi energy (particle-like configuration). In contrast to the perturbative behavior of the resonance width seen in the case of the hole-like configuration, we found that the pairing correlation works in an opposite direction in the case of the particle-like configuration, i.e. the width of the $p$-wave resonance can be reduced by the pairing.

In the present paper, we extend our study to the case of an $s$-wave quasiparticle resonance of neutron
 emerging in near-neutron-drip-line nuclei. The $s$-wave case is more interesting than the case of $l\geq 1$ since an $s$-wave resonance hardly exists in potential scattering of a neutron unless any many-body correlation takes part in. We shall discuss that the quasiparticle resonance is a promising mechanism of producing an $s$-wave resonance observable in low-energy neutron states in near-neutron-drip-line nuclei. For this purpose, we study how and in what conditions the $s$-wave resonance emerges, and we intend to reveal properties of the $s$-wave quasiparticle resonance as well as the low-energy neutron $s$-wave scattering in near-neutron-drip-line nuclei. In contrast to a pioneering work~\cite{Belyaev1987} analyzing an $s$-wave quasiparticle resonance in a system corresponding to stable nuclei, we emphasize roles of a shallow Fermi energy in weakly bound nuclei.

Similarly to the preceding study of the $p$-wave case~\cite{Kobayashi2016}, we describe unbound (resonant and non-resonant) quasiparticle states by employing the Hartree-Fock-Bogoliubov (also known as the Bogoliubov-de Genne) equation formulated in the coordinate space. Within this framework, we consider a low-energy elastic scattering of a neutron impinging on a pair-correlated target nucleus. In addition to the elastic cross section and the associated phase shift, we examine the S-matrix and its poles to explore possible resonance structures in the combined system of the target nucleus and a neutron.

Example nuclei appropriate for the study of the $s$-wave quasiparticle resonance could be neutron-rich carbon isotopes ${}^{21}$C$={}^{20}$C$+n$ or ${}^{23}$C$={}^{22}$C$+n$, which are suggested to have a neutron $s$-orbit around zero energy in addition to the features of weak binding and pairing correlation of neutrons. Indeed the drip-line isotope ${}^{22}$C is known to have a very small two-neutron separation energy less than a few hundred keV~\cite{Thoennessen2012_2,Gaudefroy2012}. A large matter rms radius~\cite{Tanaka2010,Togano2016} and a narrow momentum distribution in a two-neutron removal reaction~\cite{Kobayashi2012} suggests a two-neutron halo with a major component of $(2s_{1/2})^2$. The neighbor isotope ${}^{21}$C is unbound, and this suggests the neutron pairing correlation as a binding mechanism of ${}^{22}$C. In ${}^{20}$C, a partial occupancy of the neutron $2s_{1/2}$~\cite{Kobayashi2012,Togano2016} as well as the breaking of the $N=14$ subshell closure~\cite{Stanoiu2008} are pointed out. Thus it may be reasonable to assume the neutron pairing correlation in these isotopes. Concerning the unbound isotopes ${}^{21}$C and ${}^{23}$C, a few experimental information is available. A recent experiment~\cite{Orr2016,Nakamura2016} using one-nucleon removal reaction from ${}^{22}$C and ${}^{22}$N suggests possible existence of a low-lying $s$-wave resonance with less than 1 MeV in decay spectrum of ${}^{21}$C$\rightarrow{}^{20}$C$ +n$ whereas non-resonant $s$-wave continuum is discussed in another experiment~\cite{Mosby2013}. Theoretical investigations studying unbound states of ${}^{21}$C, in particular possibility of $s$-wave resonance, are also limited. An analysis using the deformed Woods-Saxon model suggests that a resonance is hardly observed~\cite{Hamamoto2016}. In many-body approaches such as the relativistic mean-field model~\cite{Lu2013}, the shell model~\cite{Yuan2012}, and the coupled-cluster calculation~\cite{Jansen2014}, resonance states are not discussed as unbound states are discretized in these models. Therefore it may be useful to clarify possibility of the $s$-wave resonance from the viewpoint of the pairing effect and the quasiparticle resonance. In the present work we intend to perform an exploratory study to reveal properties of the $s$-wave quasiparticle resonance in weakly bound neutron-rich nuclei, rather than to provide a precise and quantitative prediction to specific isotopes. For these reasons, we vary key model parameters such as the pairing gap, the single-particle energy (the depth of the mean-field potential) and the Fermi energy for neutrons.

This paper consists of six sections. Following this introduction, the theoretical framework and some details of numerical analysis are explained in the second and third sections, respectively. We analyze the low-energy $s$-wave scattering of neutron in terms of the elastic cross section, the phase shift, and the poles of the S-matrix. The elastic cross section and the phase shift with representative case are shown in Sect. 4. Results of systematic analysis focusing on the S-matrix poles is presented in Sect. 5. Concluding remarks are given in Sect. 6.

\section{Theoretical Framework}
\subsection{Hartree-Fock-Bogoliubov equation in coordinate space}
In order to study properties of unbound single-particle states in a pair-correlated nucleus, we formulate an elastic scattering of a neutron on an even-$N$ nucleus by means of the Hartree-Fock-Bogoliubov (HFB) theory. For simplicity the target nucleus is assumed to be spherical. The model is essentially the same as that employed in our previous study of the $p$-wave quasiparticle resonance~\cite{Kobayashi2016}. We use the notation of Refs.~\cite{Dobaczewski1984,Matsuo2001}. 

In this framework, the pairing correlation is described in terms of the pair potential $\Delta(r)$. The single-particle motion influenced by the pairing correlation is formulated as the Bogoliubov quasiparticle, whose wave function $\phi(\vec{r}, \sigma)$ has two components:
\begin{equation}
\phi(\vec{r}, \sigma)=\left(
\begin{array}{c}
\varphi_{1,lj}(r)\\
\varphi_{2,lj}(r)
\end{array}
\right)
[Y_{l}(\theta ,\varphi)\chi_{1/2}(\sigma)]_{jm}
\label{qpwf}
\end{equation}
where $\vec{r}$ and $\sigma$ are spatial coordinate and spin variable, and $Y_{l}$ and $\chi_{1/2}$ are the spherical harmonics and the spinor, respectively. $\varphi_{lj}(r)=\left( \varphi_{1,lj}(r), \varphi_{2,lj}(r) \right)^T$ are radial part of the wave function. $l$, $j$ and $m$ are the orbital and total angular-momentum quantum numbers. Isospin index is omitted for simplicity. 

The radial wave function $\varphi_{lj}(r)$ of the quasiparticle state with excitation energy $E$ obeys the radial HFB equation:
\begin{equation}
\left(
\begin{array}{cc}
h(r)-\lambda & \Delta(r)\\
\Delta(r) & -h(r)+\lambda
\end{array}
\right)\varphi_{lj}(r,E)
=E\varphi_{lj}(r,E)
\label{hfbeq}
\end{equation}
where $\lambda$ is the Fermi energy ($\lambda<0$) and $E$ is the quasiparticle energy, which is defined with respect to the Fermi energy $\lambda$. The single-particle hamiltonian $h(r)$ in Eq. (\ref{hfbeq}) is
\begin{equation}
h(r)=-\frac{\hbar^2}{2m}\frac{d^2}{dr^2}-\frac{\hbar^2}{mr}\frac{d}{dr}+\frac{\hbar^{2}l(l+1)}{2mr^{2}}+U_{lj}(r)
\end{equation}
with the nucleon mass $m$ and the mean-field potential $U_{lj}(r)$. The pair potential $\Delta(r)$ causes mixing between 
the upper component $\varphi_{1,lj}(r)$
and the lower component $\varphi_{2,lj}(r)$.

\subsection{Elastic phase shift and cross section}
The Bogoliubov quasiparticle has a character as a ``particle'' on one hand, and that of a ``hole'' on the other hand. The upper component $\varphi_{1,lj}(r)$ is the ``particle" component of the quasiparticle state while the lower component $\varphi_{2,lj}(r)$ represents a ``hole" component. If the quasiparticle state is unbound, the upper (particle) component becomes a scattering wave in the asymptotic region with the lower (hole) component vanishing asymptotically~\cite{Bulgac1980,Dobaczewski1984}. This makes it possible to describe an unbound neutron scattering on a pair-correlated even-$N$ nucleus by representing the neutron traveling outside the interaction region in terms of the particle (upper) component $\varphi_{1,lj}(r)$ in the asymptotic region $r\rightarrow \infty$ while the coupling between the upper (particle) and lower (hole) components occurs inside the nucleus due to the influence of the pair potential $\Delta(r)$.

Therefore, in order to describe the neutron elastic scattering we consider an unbound quasiparticle state, which has incoming and outgoing waves in the upper (particle) component $\varphi_{1,lj}(r)$ in the asymptotic wave function~\cite{Kobayashi2016}:
\begin{equation}
\left(
\begin{array}{c}
\varphi_{1,lj}(r,E)\\
\varphi_{2,lj}(r,E)
\end{array}
\right)_\mathrm{asympt}=
C(E)\left(
\begin{array}{c}
\cos \delta_{lj}(e)j_{l}(k_{1}r)-\sin \delta_{lj}(e)n_{l}(k_{1}r)\\
D(E)h^{(1)}_{l}(k_{2}r)
\end{array}
\right).
\label{bogoscatt}
\end{equation}
Here are the spherical Bessel function $j_{l}(k_{1}r)$, the spherical Neumann function $n_{l}(k_{1}r)$ and the phase shift $\delta_{lj}(e)$ in the upper component as well as the first-kind spherical Hankel function $h^{(1)}_{l}(k_{2}r)$ in the lower component~\cite{AbramowitzStegun}, which has an exponentially decaying form $h^{(1)}_{l}(k_{2}r) \sim \exp(-\kappa_{2}r)$ with $\kappa_{2}=\mathrm{Im}\left( k_{2}\right)=\sqrt{2m\left|\lambda-E\right|}/\hbar>0$. $C(E)$ is determined from the normalization condition:
\begin{equation}
\sum_{\sigma}\int d\vec{r}\phi^{\dagger} (\vec{r}, \sigma, E)\phi (\vec{r}, \sigma, E^{\prime})=\delta(E-E^{\prime}),
\end{equation}
giving $C(E)=\sqrt{2mk_{1}(E)/\hbar^{2}\pi}$.

The wave numbers $k_1$ and $k_2$ characterize the asymptotic behavior of $\varphi_{1,lj}(r)$ and $\varphi_{2,lj}(r)$, respectively, and they are related to the quasiparticle energy $E$ and the Fermi energy $\lambda$ as $\hbar^2 k_1^2/2m=\lambda+E$ and $\hbar^2 k_2^2/2m=\lambda-E$. A scattering (unbound) nucleon with kinetic energy $e=\hbar^2 k_1^2/2m=\lambda+E >0$ corresponds to a quasiparticle state in the energy region $E > -\lambda =|\lambda|$ whereas discrete (bound) quasiparticle states appear in the interval $0 < E < |\lambda|$. The threshold quasiparticle energy is $E_\mathrm{th}=|\lambda|$.

The radial HFB equation (\ref{hfbeq}) has two independent solutions 
$\varphi^{(1)}_{lj}(r)$ and $\varphi^{(2)}_{lj}(r)$
that are regular at the origin $r=0$;  These are easily obtained with a numerical method for differential equations. Connecting a regular solution
\begin{equation}
\varphi_{lj}(r,E)=A(E)\varphi^{(1)}_{lj}(r)+B(E)\varphi^{(2)}_{lj}(r)
\label{bogowf}
\end{equation}
to the asymptotic wave function Eq.~(\ref{bogoscatt}) at a certain radius $r=R^*$ outside the interaction region, we obtain the phase shift $\delta_{lj}(e)$ and all the other coefficients. The elastic cross section in the partial wave $lj$ is given as
\begin{eqnarray}
\sigma_{lj}(e)=\frac{4\pi}{k^{2}_{1}(E)} \left( j+\frac{1}{2} \right) \sin ^{2}\delta_{lj}(e).
\end{eqnarray}

\subsection{S-matrix in complex planes\label{subsec_smatrix}}
Analysis of the S-matrix is a powerful method to study scattering properties, in particular, resonance structures in the scattering. In the present HFB framework, the S-matrix $S_{lj}(E)$ for the elastic neutron scattering in the partial wave $lj$ is defined in terms of the asymptotic wave function, Eq.~(\ref{bogoscatt}), which is rewritten as
\begin{eqnarray}
\left(
\begin{array}{c}
\varphi_{1,lj}(r,E)\\
\varphi_{2,lj}(r,E)
\end{array}
\right)_\mathrm{asympt}
&=&
C(E)\left(
\begin{array}{c}
S_{lj}(E)h^{(1)}_{l}(k_{1}r)+h^{(2)}_{l}(k_{1}r)\\
D(E)h^{(1)}_{l}(k_{2}r)
\end{array}
\right) \nonumber \\
&\sim& \frac{1}{r}
\left(
\begin{array}{c}
\exp(-ik_{1}r)- S_{lj}(E)(-1)^l\exp(ik_{1}r)\\
D^{\prime}(E)\exp(ik_{2}r)
\end{array}
\right)
\label{smatrix}
\end{eqnarray}
where $h^{(2)}_{l}(k_{1}r)$ represents the second-kind spherical Hankel function~\cite{AbramowitzStegun}. To analyze the S-matrix we treat the energy $E$ and the wave numbers $k_{1}$ and $k_{2}$  as complex variables, which have mutual relations
\begin{equation}
k_{1}=\sqrt{\frac{2m}{\hbar^2}(\lambda+ E)},~k_{2}=\sqrt{\frac{2m}{\hbar^2}(\lambda- E)}.
\label{kpm}
\end{equation}
Since the S-matrix is specified not only by the energy $E$ but also by the asymptotic wave numbers $k_{1}$ and $k_{2}$ in Eq.~(\ref{smatrix}), the S-matrix $S_{lj}(E)$ has a four-sheeted Riemann surface, for which we introduce a branch cut on the positive real axis of $E$ with $E>|\lambda|=-\lambda$ and another branch cut on the negative real axis of $E$ with $E<-|\lambda|=\lambda$. Specifically, we relate $k_{1}$, $k_{2}$ and $E$ by
\begin{equation}
E=\frac{\hbar^2}{2m}\left| k_{1}\right|^{2}e^{2i\theta_{1}}-\lambda,~E=-\frac{\hbar^2}{2m}\left| k_{2}\right|^{2}e^{2i\theta_{2}}+\lambda
\label{kpm_theta}
\end{equation}
for $k_{i}=\left| k_{i}\right|e^{i\theta_{i}}$ $(0 \leq \theta_{i}<2\pi)$ and $i=1,2$. We term these four Riemann sheets as listed in Table.~\ref{Riemannsheets}. Note that the branch point $E=\left| \lambda\right|=-\lambda$ of the cut on the positive energy $E$-axis corresponds to the threshold for the physical scattering state. A peculiar feature is the existence of another branch point $E=-\left| \lambda\right|=\lambda$ and the associated branch cut on the negative energy $E$-axis, which results in the four-sheet structure of the Riemann surface. It originates from the quasiparticle-quasihole symmetry of the HFB equation (\ref{hfbeq}), and it brings about new aspects of the $s$-wave scattering as we will discuss in Section 5.
\begin{table}[h]
\centering
\caption{The Riemann sheets classified by $k_{1}$ and  $k_{2}$.}
\begin{tabular}{ccc} \hline
$E^{(1)}$-sheet: & $\mathrm{Im}\left(k_{1}\right)>0$, & $\mathrm{Im}\left(k_{2}\right)>0$ \\
$E^{(2)}$-sheet: & $\mathrm{Im}\left(k_{1}\right)<0$, & $\mathrm{Im}\left(k_{2}\right)>0$ \\
$E^{(3)}$-sheet: & $\mathrm{Im}\left(k_{1}\right)>0$, & $\mathrm{Im}\left(k_{2}\right)<0$ \\
$E^{(4)}$-sheet: & $\mathrm{Im}\left(k_{1}\right)<0$, & $\mathrm{Im}\left(k_{2}\right)<0$ \\ \hline
\end{tabular}
\label{Riemannsheets}
\end{table}

The S-matrix $S_{lj}(E)$ is obtained as follows. We calculate numerically two independent regular solutions, $\varphi_{lj}^{(1)}$ and $\varphi_{lj}^{(2)}$ in Eq.~(\ref{bogowf}), and connect the combined regular wave, Eq.~(\ref{bogowf}), to the asymptotic wave Eq.~(\ref{smatrix}), for given set of complex values of ($E$, $k_{1}$, $k_{2}$) which satisfy Eq.~(\ref{kpm_theta}), or equivalently, for given point on the four-sheeted Riemann surface. Four connection relations for the upper and lower components and their derivatives of the wave functions determine coefficients $A(E)$, $B(E)$, and $D(E)$ as well as the S-matrix $S_{lj} (E)$.

It is noted that the HFB equation (\ref{hfbeq}) has a structure of coupled-channel equations for $\varphi_{1,lj}$ and $\varphi_{2,lj}$. It may be possible to introduce a two-channel S-matrix
\cite{Mizuyama2019}\footnote{An attempt toward this direction is made recently in Ref.~\cite{Mizuyama2019}, in which, however, incorrect Riemann surface of the S-matrix appears to be assumed.}.
In such a formulation, $S_{lj}(E)$ defined in Eq.~(\ref{smatrix}) corresponds to a diagonal component for the upper channel $\varphi_{1,lj}$ in the $2\times 2$ S-matrix while $D(E)$ in Eq.~(\ref{smatrix}) is related to an off-diagonal component. Note however that the diagonal component is sufficient in the physical region where the second channel is closed.

\section{Details of model calculation}
In the following, we investigate the elastic $s$-wave scattering of a low-energy neutron on a weakly bound neutron-rich nucleus which has a neutron $2s_{1/2}$ single-particle orbit around zero energy. The target nucleus is supposed to represent ${}^{20}$C or ${}^{22}$C, and the total system is then ${}^{21}$C$={}^{20}$C$ +n$ or ${}^{23}$C$={}^{22}$C$+n$ with spin and parity $J^\pi=1/2^+$. However, we do not make a fine tuning of the model setting to describe a specific system, but rather we vary some key parameters of the model in a flexible way in order to perform exploratory studies to reveal possible features of unbound quasiparticle states and quasiparticle resonance in the $s$-wave.

We employ the Woods-Saxon model as the mean-field potential $U_{lj}(r)$ for neutrons:
\begin{equation}
U_{lj}(r)=\left[ V_{0}+ (\vec{l}\cdot\vec{s})V_{\mathrm{SO}}\frac{r^{2}_{0}}{r}\frac{d}{dr}  \right] f(r)
\end{equation}
\begin{equation}
f(r)=\left[ 1+\exp \left( \frac{r-R}{a_{\mathrm{d}}} \right) \right]^{-1},
\end{equation}
where the parameters $V_{0}$, $V_{\mathrm{SO}}$, and $a_{\mathrm{d}}$ are taken from  Ref.~\cite{BohrMottelson}. $R$ is the nuclear radius $R=r_{0}A^{1/3}$ with mass number $A$, which is $R=3.477$ fm for ${}^{20}$C. For the pair potential $\Delta(r)$ we adopt also a local potential of the Woods-Saxon form for simplicity:
\begin{equation}
\Delta(r)=\Delta_{0}f(r).
\end{equation}
The strength of pair potential $\Delta_{0}$ is controlled by the average pairing gap $\bar{\Delta}$~\cite{Hamamoto2003,Hamamoto2004}:
\begin{equation}
\bar{\Delta}=\int^{\infty}_{0}r^{2}\Delta (r)f(r)dr\left/\int^{\infty}_{0}r^{2}f(r)dr\right. .
\end{equation}

The neutron pairing gap is one of the key parameters in the present study. The average pairing gap $\bar{\Delta}$ of the neutron pair potential is varied in an interval $\bar{\Delta}=0 - 3$ MeV since the empirical systematics is $\Delta^{(\mathrm{exp})}\approx 12/\sqrt{A}\approx 2.7$ MeV. Another key parameter is the single-particle energy of the neutron $2s_{1/2}$ orbit. This is controlled by varying the depth $V_0$ of the Woods-Saxon potential $U_{lj}(r)$. We shift $V_0$ by $\Delta V_0$ so that the model covers situations where the neutron $2s_{1/2}$ orbit is both bound and unbound. Table~\ref{spene_20C_sdatble} shows the energy of bound states, or the energy and the full width at half maximum (FWHM) of a resonance, for neutron single-particle states $2s_{1/2}$, $1d_{5/2}$ and $1d_{3/2}$ obtained with $\Delta V_0=$0.0, 4.0 and 8.0 MeV and without the pair potential. Note that in the case of $\Delta V_0=$4.0 MeV the $2s_{1/2}$ orbit is bound only very weakly $e_{2s_{1/2}}=-0.242$ MeV. In this case we put the neutron Fermi energy at $\lambda=-0.230$ MeV which is chosen between $2s_{1/2}$ and $1d_{5/2}$. This setting is adopted as a representative case since this extreme case is useful to elucidate the effect of the weak binding. In the case of $\Delta V_0=$0.0 MeV the $2s_{1/2}$ orbit is slightly more bound ($e_{2s_{1/2}}=-1.115$ MeV), but still the binding is weak. For $\Delta V_0=$8.0 MeV, the $2s_{1/2}$ orbit is unbound, and in fact it is a virtual state. For systematic studies, we vary $\Delta V_0$ within this interval, which approximately covers the range of neutron separation energy in $^{20-22}$C.
\begin{table}[h]
\centering
\caption{The single-particle energy of neutron orbits in the Woods-Saxon potential with the parameters for $^{20}$C~\cite{BohrMottelson}, but with a variation in the potential depth $\Delta V_{0}=0.0, ~4.0$ and 8.0 MeV. The resonance energy and the resonance width (shown in the bracket) evaluated from the S-matrix pole are listed for the $1d_{3/2}$ and $1d_{5/2}$ orbits. The $2s_{1/2}$ orbit for $\Delta V_{0}=8.0$ MeV is unbound and it is a virtual state. }
\begin{tabular}{ccccccc} \hline
$\Delta V_{0}$ [MeV]  && 0.0 && 4.0 && 8.0\\ \hline
$e_{2s_{1/2}}$ [MeV] && $-1.115$ && $-0.242$ && $-$\\
$e_{1d_{3/2}}$ [MeV] && $2.174~(0.945)$ && $3.255~(2.427)$ && $4.165~(4.597)$ \\
$e_{1d_{5/2}}$ [MeV] && $-1.882$ && $-0.221$ && $1.074~(0.217)$ \\ \hline
\end{tabular}
\label{spene_20C_sdatble}
\end{table}

The regular radial wave function (\ref{bogowf}) is obtained by solving the HFB equation (\ref{hfbeq}) from the origin using the 4-th order Runge-Kutta method and a radial step $\Delta r= 0.02$ fm. We calculate the phase shift $\delta_{lj}(e)$ and the cross section $\sigma_{lj}(e)$ in the physical region $e>0$, i.e. for real values of $E>\left| \lambda\right|$. We use the same numerical method to evaluate the S-matrix, but in this case we use the complex numbers for $E$, $k_{1}$ and $k_{2}$ as discussed in Sect. \ref{subsec_smatrix}. We use the simplex method to search minimum points (zeros) of $\left|S_{lj}(E)\right |^{-2}$, i.e., poles of the S-matrix.

\section{Pairing effects on the s-wave elastic scattering}
\subsection{Elastic cross section and phase shift}
\begin{figure}[h]
\begin{minipage}{0.48\hsize}
\includegraphics[width=77mm]{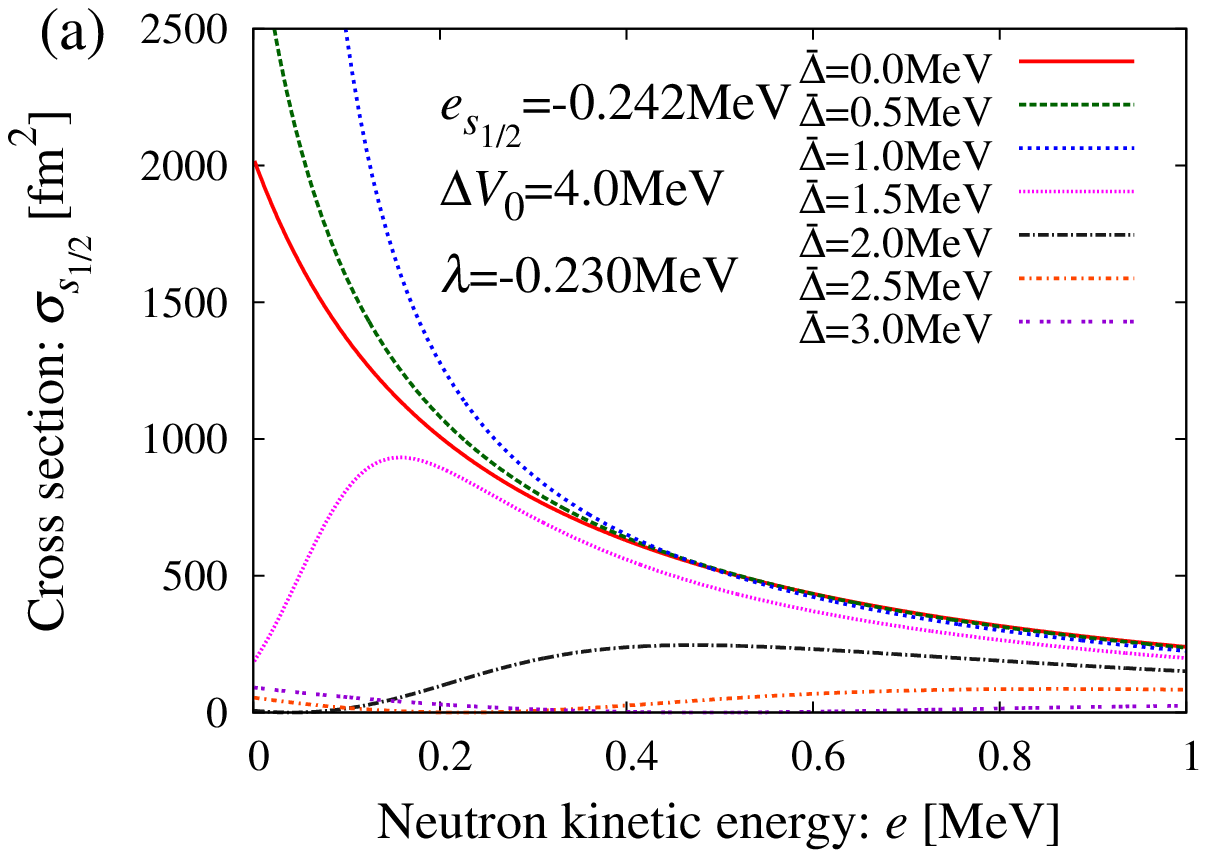}
\end{minipage}
\begin{minipage}{0.48\hsize}
\includegraphics[width=77mm]{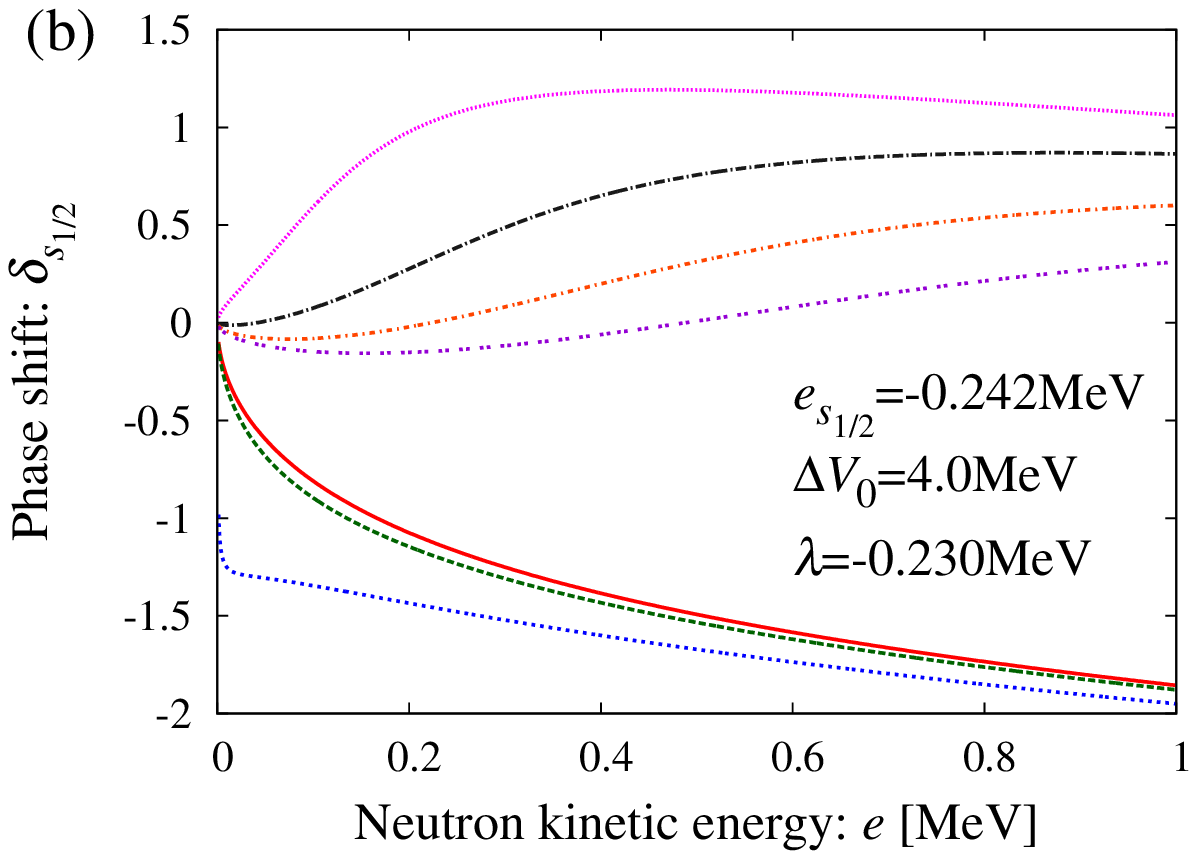}
\end{minipage}
\caption{(a) Elastic cross section $\sigma_{s_{1/2}}(e)$ and (b) phase shift $\delta_{s_{1/2}}(e)$ in the partial wave $s_{1/2}$ for various values of the average pairing gap $\bar{\Delta}$. The horizontal axis is the kinetic energy of incident neutron $e$ which is related to the quasipartcile energy $E$ with $e=\lambda+E$. The single-particle energy of the neutron $2s_{1/2}$ orbit is $e_{2s_{1/2}}=-0.242$ MeV with the potential shift $\Delta V_0=4.0$ MeV, and the Fermi energy is $\lambda=-0.230$ MeV.}
\label{fig1}
\end{figure}

Here, we focus on the representative case of very weak binding, i.e., $e_{2s_{1/2}}=-0.242$ MeV with $\Delta V_0=4.0$ MeV, and the Fermi energy at $\lambda=-0.230$ MeV. We calculate the phase shift and the cross section for the low-energy elastic scatting with different values of the average pairing gap, varied from $\bar{\Delta}=0$ to 3 MeV. The results are shown as a function of the kinetic energy of incident neutron $e=E+\lambda=E-|\lambda|$ in Fig.~\ref{fig1}.

It is seen that the pairing correlation causes strong influence on both the elastic cross section $\sigma_{s_{1/2}}(e)$ and the phase shift $\delta_{s_{1/2}}(e)$. With vanishing pair potential $\bar{\Delta}=0$, the cross section $\sigma_{s_{1/2}}(e)$ exhibits a monotonically decreasing behavior, which is typical for the $s$-wave potential scattering influenced by a bound state located just below the threshold (in fact, $e_{2s_{1/2}}=-0.242$ MeV). For a slightly increased $\bar{\Delta}=1.0 $ MeV, we see an almost diverging cross section at $e=0$, i.e., suggesting influence of a possible virtual state. With slightly increased $\bar{\Delta}=1.5$ MeV, the cross section and the phase shift behave differently; the cross section shows a maximum around $e=0.16$ MeV, which points to possibility of a low-energy resonance. As $\bar{\Delta}$ exceeds 2 MeV, the distinct structures tend to vanish.

\subsection{Analysis using the effective range}
We analyze the above result by means of the effective range expansion of the phase shift
\begin{equation}
k\cot\delta(k)=-\frac{1}{a_{\mathrm{s}}}+\frac{1}{2}k^{2}r_{\mathrm{eff}}+{\cal O}(k^4)
\label{effrangeformula}
\end{equation}
with the scattering length $a_{\mathrm{s}}$ and effective range $r_{\mathrm{eff}}$, which characterize the low-energy $s$-wave scattering~\cite{Messiah,Taylor,Kukulin,Sitenko}. To do so, we rewrite Eq.~(\ref{effrangeformula}) using the asymptotic wave number $k_{1}$ of the scattering wave $\varphi_{1,lj}(r)$ as  
\begin{equation}
\delta_{s_{1/2}}(k_{1})=\arctan\left[ k_{1}\left(-\frac{1}{a_{\mathrm{s}}}+\frac{1}{2}r_{\mathrm{eff}}k^{2}_{1} \right)^{-1}\right],
\label{effrangeformula2}
\end{equation}
and we fit it to the numerically calculated phase shift $\delta_{s_{1/2}}$ (Fig.~\ref{fig1} (b)) to obtain the scattering length $a_{\mathrm{s}}$ and the effective range $r_{\mathrm{eff}}$. The range of fitting is $0 <e<0.1$ MeV, which satisfies the applicability condition $kr_{\mathrm{eff}}\ll1.0$~\cite{Messiah} for the effective range expansion Eq.~(\ref{effrangeformula}), with $r_{\mathrm{eff}}\sim 5$ fm estimated from the potential radius.

The obtained scattering length $a_{\mathrm{s}}$ and effective range $r_{\mathrm{eff}}$ are listed in Table~\ref{areff} for various values of the average pairing gap $\bar{\Delta}$. It is seen that both $a_{\mathrm{s}}$ and $r_{\mathrm{eff}}$ are influenced strongly by the pairing correlation, as is easily expected from the peculiar behavior of the phase shift shown in Fig.~\ref{fig1} (b). A remarkable feature is that the scattering length $a_{\mathrm{s}}$ becomes negative values around $\bar{\Delta} \approx 1.5-2.0$ MeV and that the effective range $r_{\mathrm{eff}}$ also becomes negative for $\bar{\Delta} \approx 1.0-3.0$ MeV (and even very large negative values $-100 \sim -150$ fm). Such a behavior is not expected from a simple potential scattering in the $s$-wave, for which the effective range is expected to be a positive value with the order of potential size $\sim 5$ fm, and the scattering length behaves smoothly. More specifically its inverse $1/a_{\mathrm{s}}$ varies monotonically with small change in a potential parameter. In contrast, a negative effective range is known to realize if there exists a Feshbach resonance~\cite{Blackley2014} or a resonance near the threshold~\cite{Hyodo2013,Ikeda2011}. The result shown in Table.~\ref{areff} suggests non-simple behavior of the elastic scattering and also possibility of the low-energy $s$-wave resonance under the influence of the pairing correlation.
\begin{table}[h]
\centering
\caption{The scattering length $a_{\mathrm{s}}$ and effective range $r_{\mathrm{eff}}$ extracted from the $s_{1/2}$ phase shift $\delta_{s_{1/2}}$ for various values of the average pairing gap $\bar{\Delta}$. The single-particle energy of the neutron $e_{2s_{1/2}}=-0.242$ MeV and the Fermi energy $\lambda=-0.230$ MeV are the same as those in Fig.~\ref{fig1}.}
\begin{tabular}{ccc} \hline
$\bar{\Delta}$ [MeV] & $a_{\mathrm{s}}$ [fm] & $r_{\mathrm{eff}}$ [fm]\\ \hline
0.0 & 12.658 & 5.373\\
0.5 & 15.385 & 3.831\\
1.0 & 121.212 & $-1.478$\\
1.5 & $-4.823$ & $-45.341$\\
2.0 & $-1.078$ & $-109.617$\\
2.5 & 2.276 & $-156.011$\\
3.0 & 3.165 & $-69.521$\\ \hline
\end{tabular}
\label{areff}
\end{table}

\section{S-matrix poles and quasiparticle resonance}
\subsection{Four poles}
To reveal the origin of the peculiar behavior of the pairing effect, we analyze the S-matrix with focus on its poles. We first choose the case of $\bar{\Delta}=1.5$ MeV as a representative case, and then study dependence on $\bar{\Delta}$.

\begin{figure}[h]
\begin{center}
\includegraphics[width=90mm]{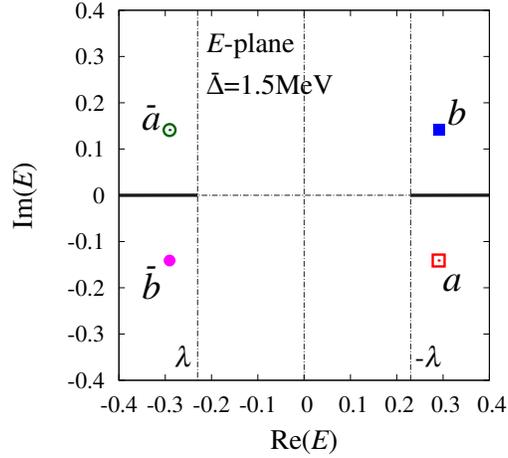}
\end{center}
\caption{Position of the four poles ($a, \bar{a}, b$, and $\bar{b}$) for $\bar{\Delta}=1.5$ MeV in the complex $E$-plane with $e_{2s_{1/2}}=-0.242$ MeV ($\Delta V_0=4.0$ MeV) and $\lambda=-0.230$ MeV. The poles $a$ and $b$ are located on the second Riemann sheet $E^{(2)}$ while $\bar{a}$ and $\bar{b}$ are on the third Riemann sheet $E^{(3)}$. Two thick lines on the real $E$-axis are the branch cuts.}
\label{fig2}
\end{figure}

\begin{figure}[h]
\begin{minipage}{0.48\hsize}
\includegraphics[width=90mm]{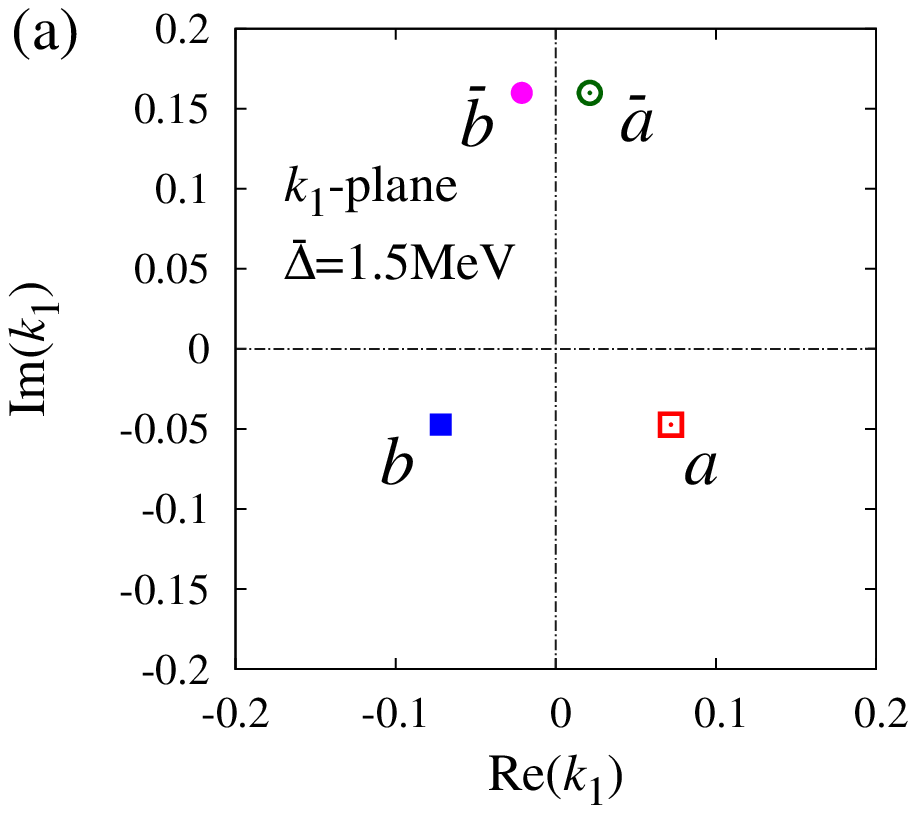}
\end{minipage}
\begin{minipage}{0.48\hsize}
\includegraphics[width=90mm]{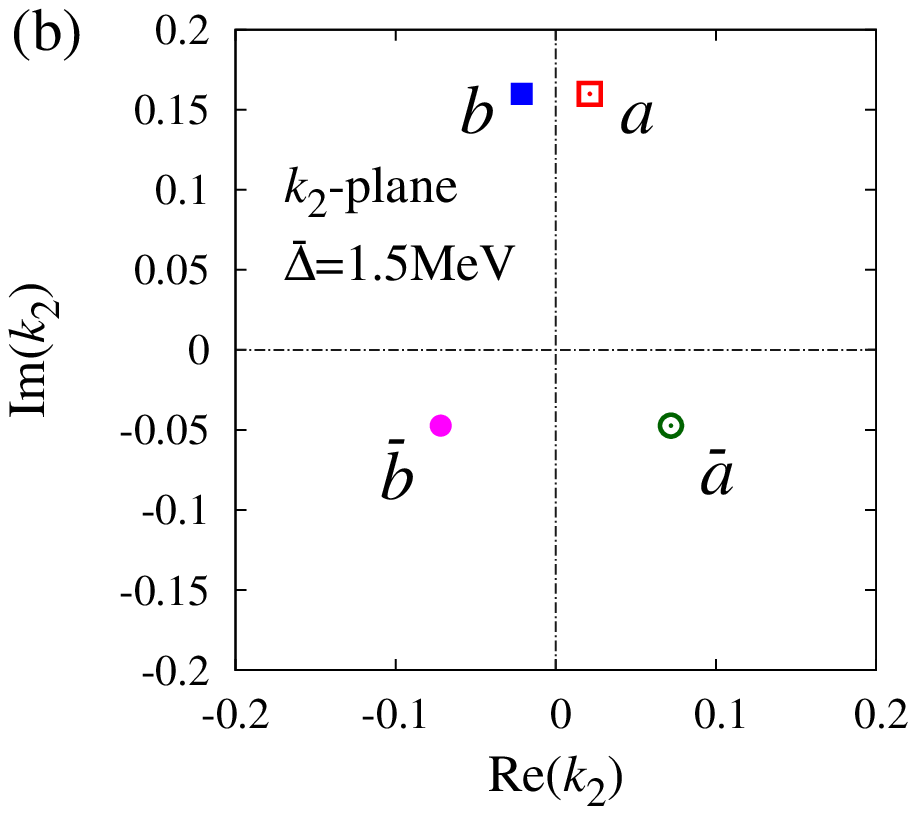}
\end{minipage}
\caption{The same as Fig.~\ref{fig2}, but in the complex (a) $k_{1}$- and  (b) $k_{2}$-planes. } 
\label{fig3}
\end{figure}

We search poles on all of the four Riemann sheets of the complex $E$-plane, and found four poles shown in Fig.~\ref{fig2}. The positions of these poles are also shown in the complex $k_{1}$- and $k_{2}$-planes in Fig.~\ref{fig3}. To these poles we assign labels $a, \bar{a}, b$, and $\bar{b}$ with a naming rule explained just below.

We note here that the HFB equation (\ref{hfbeq}) has a symmetry with respect to an exchange of the upper (particle) and lower (hole) components of the wave function, called the quasiparticle-quasihole symmetry~\cite{RingSchuck,Dobaczewski2013}:
\begin{equation}
\left(
\begin{array}{c}
\varphi_{1,lj}(r,-E)\\
\varphi_{2,lj}(r,-E)
\end{array}
\right)
=
\left(
\begin{array}{c}
-\varphi_{2,lj}(r,E)\\
\varphi_{1,lj}(r,E)
\end{array}
\right)
\label{qpqhsym}
\end{equation}
which accompanies the sign change of the quasiparticle energy $E$. Since the wave functions of the poles also have this symmetry, the four poles are related pairwise as $a$-$\bar{a}$ (and $b$-$\bar{b}$), where $a$ and $\bar{a}$ ($b$ and $\bar{b}$ also) are mutually conjugate with respect to the quasiparticle-quasihole symmetry. Consequently, the positions $(k_{1},k_{2},E)_{a}$ and $(k_{1},k_{2},E)_{\bar{a}}$ of the poles $a$ and $\bar{a}$ are related to each other as
\begin{equation}
(k_{1},k_{2},E)_{\bar{a}} = (k_{2},k_{1},-E)_a ,
\end{equation}
and the same applies to the poles $b$ and $\bar{b}$. We note also the Schwartz reflection principle~\cite{Taylor,Kukulin,Sitenko}, which relates a pair of poles at $(k_{1},k_{2},E)$ and $(-k_{1}^\ast,-k_{2}^\ast,E^\ast)$ unless $k_{1,}$ and $k_{2}$ are pure imaginary and $E$ is real. This relation holds between $a$ and $b$ as well as $\bar{a}$ and $\bar{b}$ in the case of Figs.~\ref{fig2} and~\ref{fig3}. Among the four poles, we assigned label $a$ to the one which is located in the fourth quadrant of the complex $k_{1}$-plane, i.e. the lower half of the second sheet $E^{(2)}$. This pole can be interpreted as a {\it resonance pole} since the wave function of the upper component $\varphi_{1,lj}(r)$ is outgoing and exponentially growing with $r$ (as $\mathrm{Re}\left(k_{1}\right)>0$ and $\mathrm{Im}\left(k_{1}\right)<0$), and it is expected to cause a resonance structure in observables. We assigned the label $b$ to the counter part of $a$ with respect to the Schwartz reflection principle, which is located in the third quadrant of the complex $k_{1}$-plane, and in the upper half of the  $E^{(2)}$-sheet. It is often called an anti-resonance pole. The other poles $\bar{a}$ and $\bar{b}$ are not resonance poles since the associated wave function $\varphi_{1,lj}(r)$ is exponentially decreasing with $r$ ($\mathrm{Im}\left(k_{1}\right)>0$). We will discuss roles of these poles later.

It happens in other cases that poles are located on the real axis of the $E$-plane. An example is the case of $\bar{\Delta}=0.5$ MeV, where we found two poles located pairwise at $E=E_0$ and $E=-E_0$ ($0<E_0<|\lambda|$) on the real $E$-axis in the $E^{(1)}$-sheet (see Fig.~\ref{fig4}). These are physical discrete eigenstates of the HFB equation, and the pole at $E=E_0$ corresponds to a bound quasiparticle state with positive quasiparticle energy $E=E_0$ while the negative energy pole at $E=-E_0$ is the quasihole counterpart of the positive energy quasiparticle state. We assign the labels $a$ and $\bar{a}$ to these states. In the $k_{1}$- and $k_{2}$-planes, these poles are located on the positive imaginary axis, and hence the wave function is exponentially decaying at large $r$, i.e. the property of physical bound states. There exist other two poles, for which one of $k_1$ and $k_2$ is located on the negative imaginary axes, and hence on the unphysical sheets $E^{(2)}$ and $E^{(3)}$.  We assign label $b$ and $\bar{b}$ to the poles on the $E^{(2)}$- and $E^{(3)}$-sheet, respectively. This rule of labeling is consistent with that discussed for the case of Figs.~\ref{fig2} and~\ref{fig3} in a sense that the same label is kept along a pole trajectory when a pole moves continuously (see Fig.~\ref{fig4}).

\subsection{Pairing effect and pole trajectories}\label{subsec:trajectory}
\begin{figure}[h]
\begin{minipage}{0.48\hsize}
\includegraphics[width=90mm]{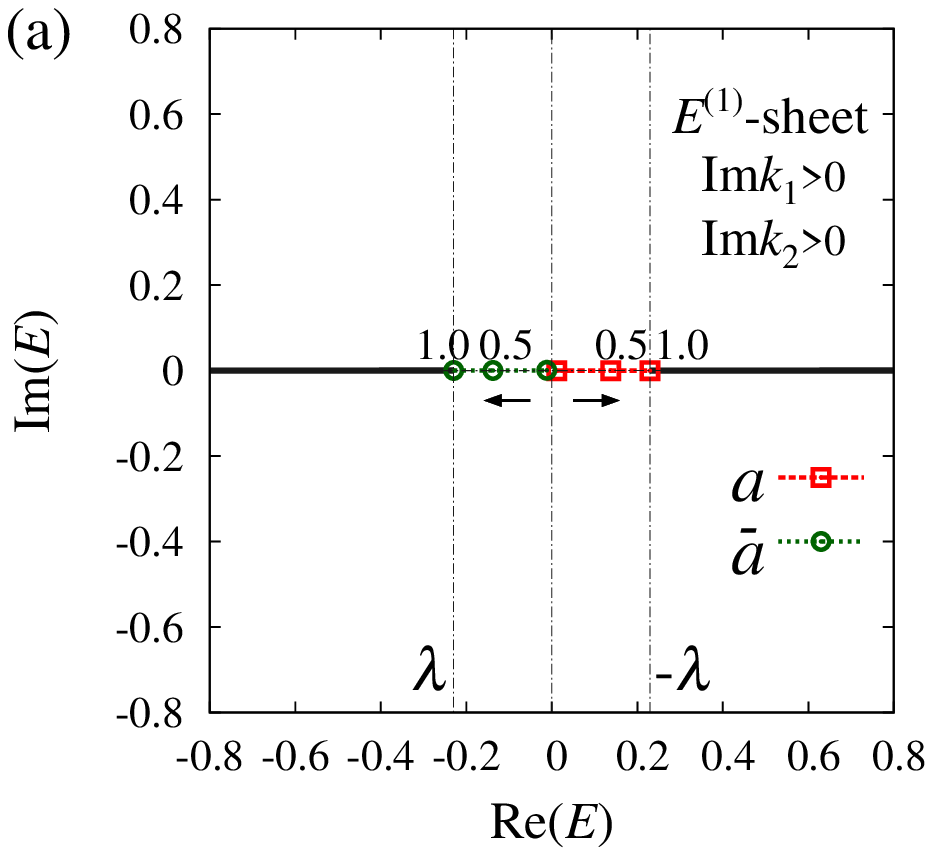}
\end{minipage}
\begin{minipage}{0.50\hsize}
\includegraphics[width=90mm]{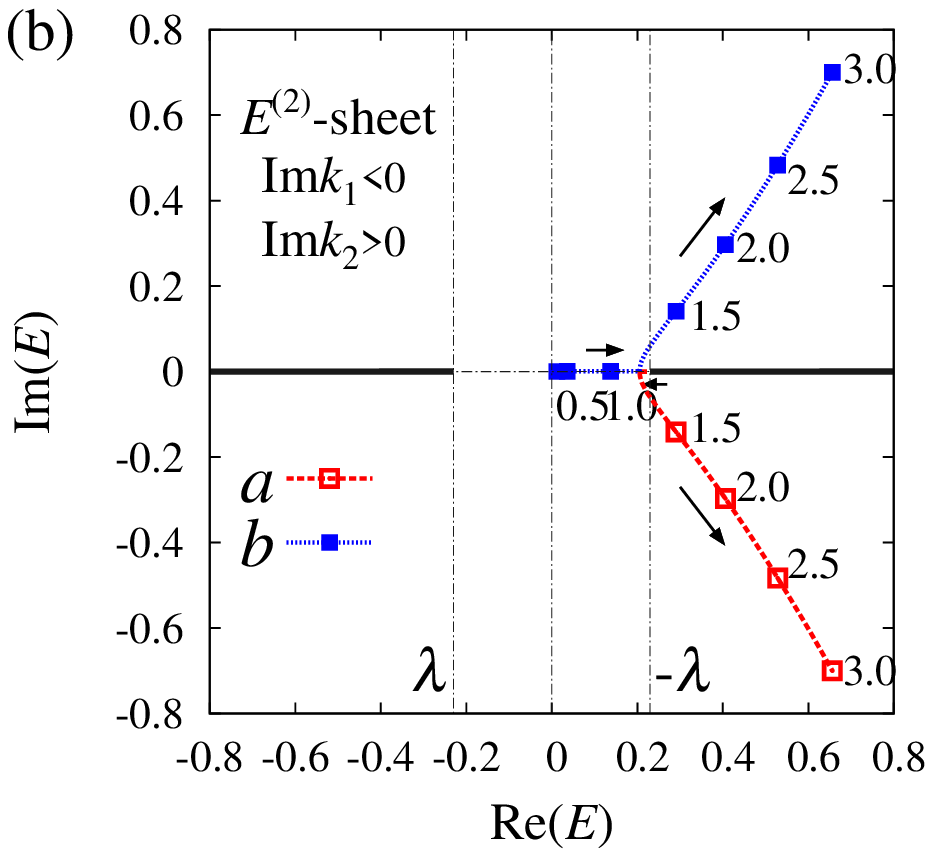}
\end{minipage}
\begin{minipage}{0.48\hsize}
\includegraphics[width=90mm]{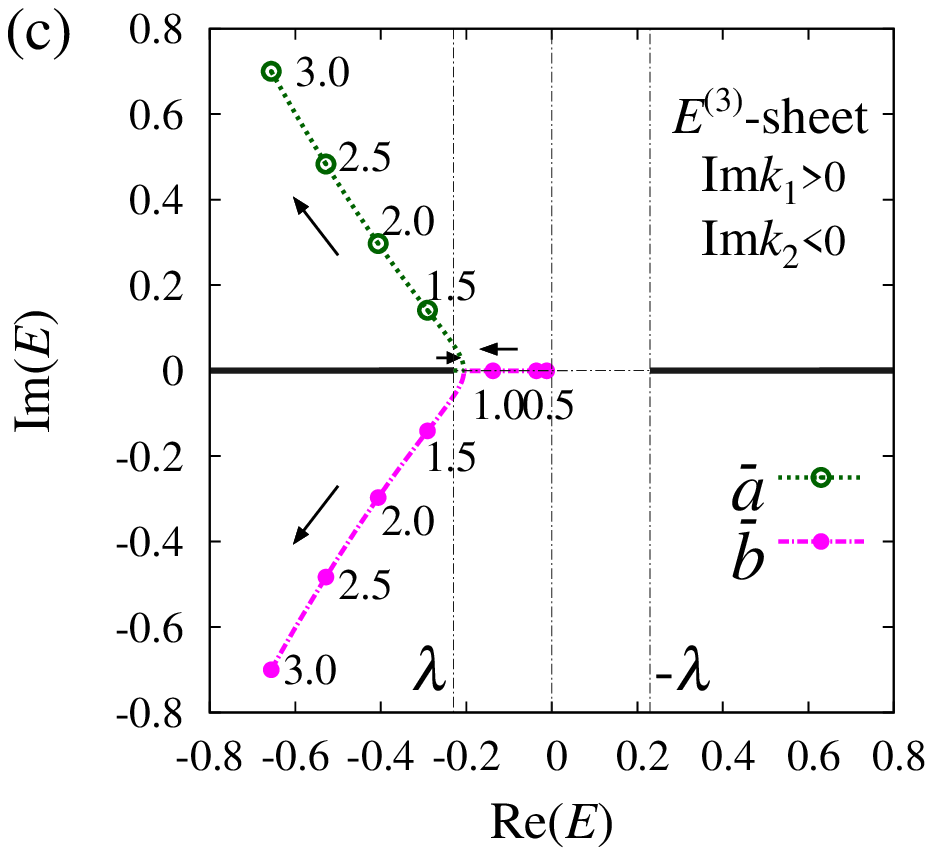}
\end{minipage}
\begin{minipage}{0.48\hsize}
\includegraphics[width=90mm]{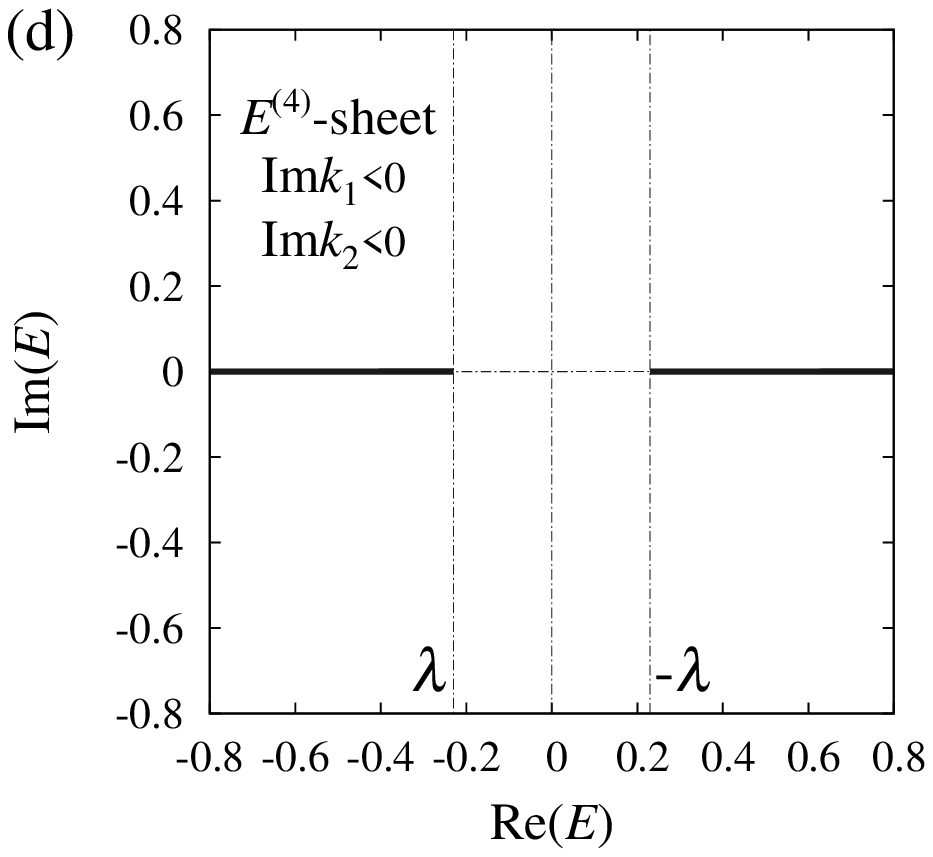}
\end{minipage}
\caption{Pole trajectories on the Riemann sheets of the complex $E$-plane, where the average pairing gap
$\bar{\Delta}$ is varied in the range $\bar{\Delta}=0 - 3$ MeV while the $2s_{1/2}$ single-particle energy and the Fermi energy are fixed to $e_{2s_{1/2}}=-0.242$ MeV ($\Delta V_0=4.0$ MeV) and $\lambda=-0.230$ MeV. Trajectories are plotted separately on the four Riemann sheets, and marked with symbols for every 0.5 MeV intervals in $\bar{\Delta}$. Two thick lines on the real $E$-axis are the branch cuts.}
\label{fig4}
\end{figure}

\begin{figure}[h]
\begin{minipage}{0.48\hsize}
\includegraphics[width=90mm]{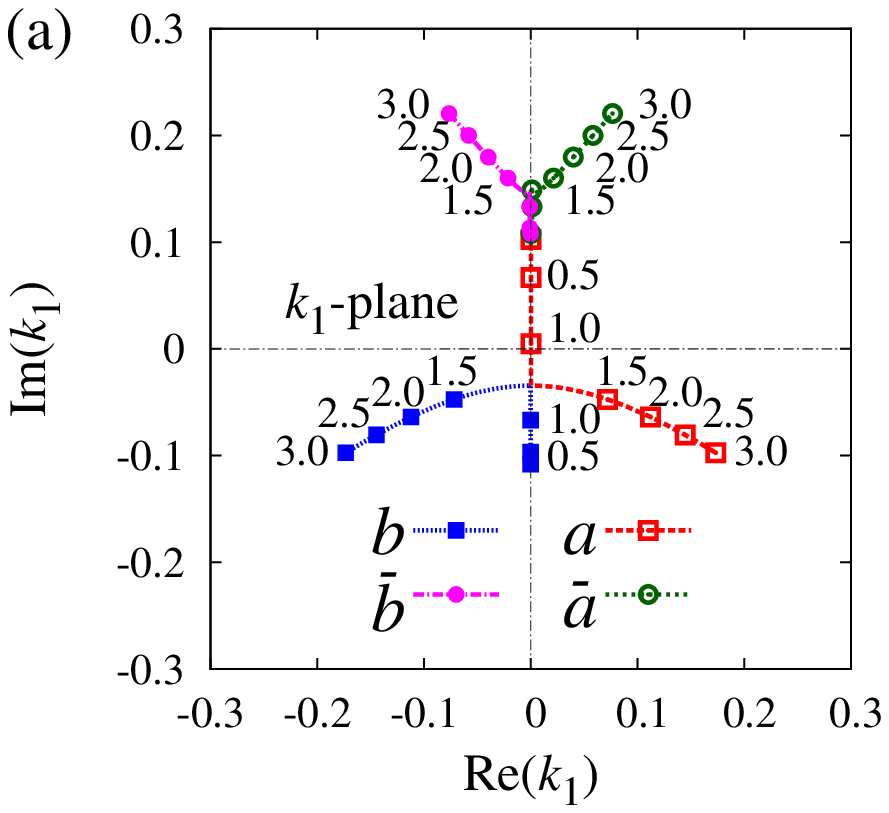}
\end{minipage}
\begin{minipage}{0.48\hsize}
\includegraphics[width=90mm]{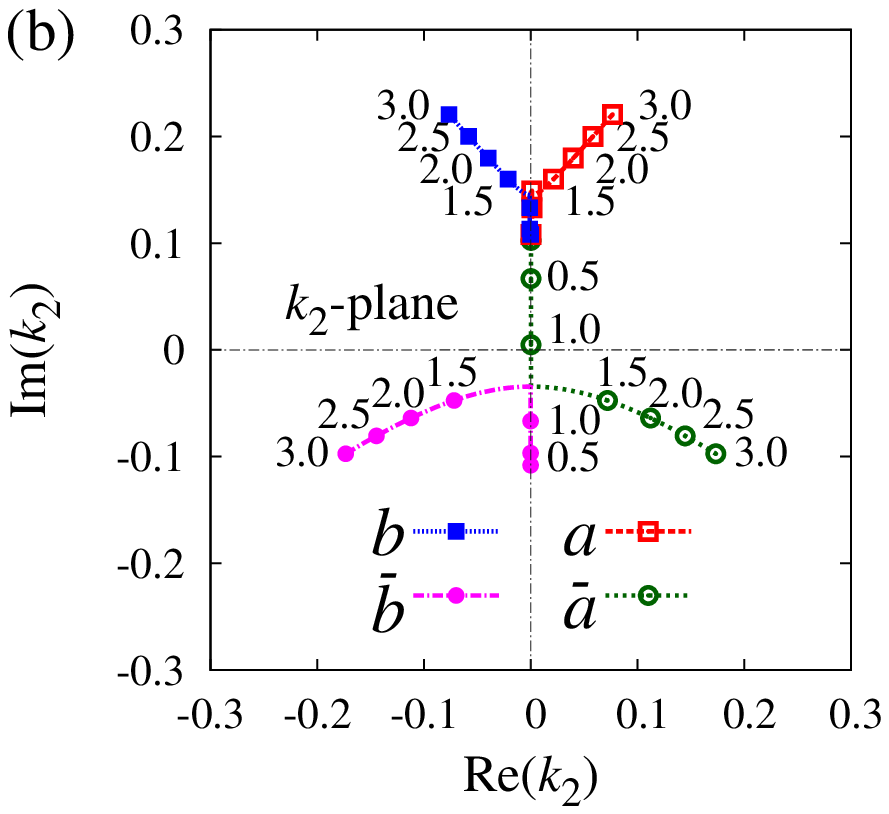}
\end{minipage}
\caption{The same as Fig.~\ref{fig4} but in the complex (a) $k_{1}$- and (b) $k_2$-planes.}
\label{fig5}
\end{figure}

Figure~\ref{fig4} shows trajectories of the four poles on the Riemann sheets of the $E$-plane as a function of the average pairing gap $\bar{\Delta}$, which is varied from $\bar{\Delta}=0$ to $\bar{\Delta}=3.0$ MeV. The trajectories on the complex $k_1$- and $k_2$-planes are also shown in Fig.~\ref{fig5}. These figures illustrate how the poles move along the trajectories with the increase of the average pairing gap $\bar{\Delta}$. We mainly discuss the positions in the $E$-plane and in the $k_1$-plane. The positions of $a$, $b$, $\bar{a}$, and $\bar{b}$ in the $k_2$-plane are related to those in the $k_1$-plane via $k_{2a}=k_{1\bar{a}}, k_{2\bar{a}}=k_{1a},k_{2b}=k_{1\bar{b}}$, and $k_{2\bar{b}}=k_{1b}$.

Let us fist discuss the limit of $\bar{\Delta}\rightarrow 0$. In this limit the poles $a$ and $\bar{a}$ are located on the real $E$ axis of the physical $E^{(1)}$-sheet at $E_a=|e_{2s_{1/2}}-\lambda|$ and $E_{\bar{a}}=-|e_{2s_{1/2}}-\lambda|$ with $ |e_{2s_{1/2}}-\lambda|=0.012$ MeV. These are physical bound quasiparticle states, conjugate to each other with respect to the quasiparticle-quasihole symmetry. The pole positions in the complex $k_1$- and $k_2$-planes are $(k_1,k_2)_{a}=(i\kappa_{+},i\kappa_{-})$ and $(k_1,k_2)_{\bar{a}}=(i\kappa_{-},i\kappa_{+})$ with $\kappa_{+}=\sqrt{2m(-E_a-\lambda)}/\hbar =0.108$ fm$^{-1}$ $\kappa_{-}=\sqrt{2m(E_a-\lambda)}/\hbar =0.103$ fm$^{-1}$. The other poles $b$ and $\bar{b}$ have the same energies $E_b=|e_{2s_{1/2}}-\lambda|$ and $E_{\bar{b}}=-|e_{2s_{1/2}}-\lambda|$ as $E_a$ and $E_{\bar{a}}$. However, they lies on the second and the third Riemann sheets, respectively, i.e. on the lower-half $k_1$- and $k_2$-planes: $(k_1,k_2)_{b}=(-i\kappa_{+},i\kappa_{-})$ and $(k_1,k_2)_{\bar{b}}=(i\kappa_{-}, -i\kappa_{+})$.

\begin{figure}[h]
\begin{center}
\includegraphics[width=120mm]{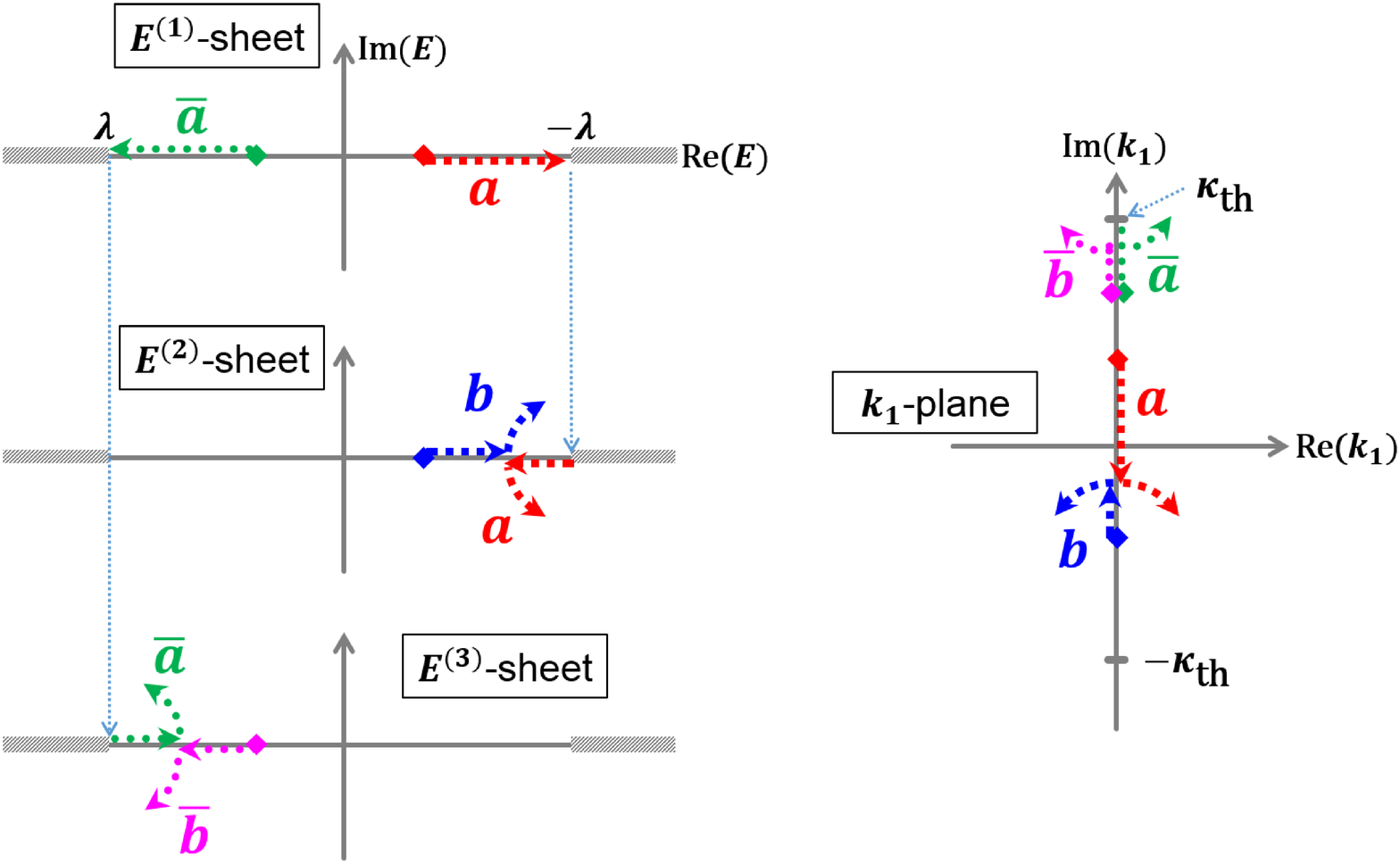}
\end{center}
\caption{Schematic illustration of the pole trajectory for small $\bar{\Delta}$ on the Riemann sheets of the complex $E$-plane and in the complex $k_{1}$-planes. Two gray thick lines on the real $E$-axis are the branch cuts.}
\label{fig6}
\end{figure}

As $\bar{\Delta}$ increases slightly ($\bar{\Delta}=0.5$ MeV, for example), these four poles move along the real $E$ axis on respective Riemann sheets, but with different energies $E_{a}\neq E_{b}$. We show in Fig.~\ref{fig6} a schematic illustration of the pole trajectory at small $\bar{\Delta}$ since the four poles with small $\bar{\Delta}$ are hardly distinguished in Figs.~\ref{fig4} and~\ref{fig5} because of the small number $|e_{2s_{1/2}}-\lambda|=0.012$ MeV in the present parameter set.

Increasing $\bar{\Delta}$ further, the poles $a$ and $b$ move toward the threshold, i.e. the branch point $E=-\lambda, k_1=0, k_2=i\kappa_{\mathrm{th}}$ ($E=-\lambda=|\lambda|$ is the threshold energy for becoming unbound, and $\kappa_{\mathrm{th}}=\sqrt{4m|\lambda|}/\hbar$). With $\bar{\Delta}$ slightly below 1.0 MeV, the pole $a$ reaches the branch point, and then moves from the first to the second Riemann sheet. In the $k_1$-plane, the pole $a$ moves downward along the positive imaginary axis, and after passing the origin $k_1=0$ (corresponding to the threshold) it moves further downward along the negative imaginary axis. This behavior of $a$ is typical of a virtual state. The trajectory of the pole $b$ is different from that of $a$. For small $\bar{\Delta}$, $b$ in the $k_1$-plane moves upward along the negative imaginary axis. Proceeding further, the poles $a$ and $b$ merge at $k_1=-0.034 \times i $ fm$^{-1}$  with $\bar{\Delta}=1.106$ MeV, and they leave the imaginary axis. Then the pole $a$ moves in the fourth quadrant of the $k_1$-plane (and in the lower-half $E$-plane on the second Riemann sheet). It is nothing but a resonance pole. The pole $b$ moves in the third quadrant in the $k_1$-plane (the upper-half $E$-plane of the second sheet). It is an anti-resonance pole and a Schwartz reflection of $a$. This situation is what we already discussed with Figs.~\ref{fig2} and \ref{fig3}.

Regarding the pole trajectories shown in Figs.~\ref{fig4} and \ref{fig5}, we remark three distinct features; i) A low-energy resonance pole  with a moderate width emerges. ii) The four poles are located at close positions near the threshold. iii) The four poles are grouped as two pairs.

i) \underline{Emergence of a low-energy resonance pole with moderate width:} The real and imaginary parts of the resonance pole $a$ in the complex $E$-plane are related to the resonance energy $e_{\mathrm{res}}=\mathrm{Re}\left(E_{a}\right)-|\lambda|$ and the resonance width $\Gamma_{\mathrm{res}}=-2\mathrm{Im}\left(E_a\right)$. As seen in Fig.~\ref{fig4}, the resonance width $\Gamma_{\mathrm{res}}$ is comparable to $e_{\mathrm{res}}$ for $\bar{\Delta}\approx 1.5 \sim 2.0$ MeV, hence a resonance structure can be expected in physical observables; Indeed a peak at $e\approx 0.16$ MeV appears in the elastic cross section in the case of $\bar{\Delta}=1.5$ MeV (Fig.~\ref{fig1} (a)). If we consider the $s$-wave scattering caused only by the mean-field potential, i.e. if the pair potential is set to zero, we do not have a resonance pole with a narrow or moderate resonance width irrespective of the potential depth. 
We emphasize that a crucial difference from the potential scattering is the presence of the pole $b$. The wave function of the pole $b$ has a bound state structure in the hole component $\varphi_{2}(r)$ (whose asymptotics is exponentially decaying  $\sim e^{-\mathrm{Im}\left(k_{2,b}\right) r}$ with $\mathrm{Im}\left(k_{2,b}\right)>0$) while the particle component $\varphi_{1}(r)$  is exponentially diverging with $\mathrm{Im}k_{1,b}<0$. Such a unphysical state is allowed to exist only when the pair potential $\Delta(r)$ causes a coupling between the scattering wave $\varphi_{1}(r)$ and the bound hole component $\varphi_{2}(r)$. Furthermore, we note that in the present situation shown in Figs.~\ref{fig4} and \ref{fig5}, the pole $b$ merges with the pole $a$ at a position close to the threshold, and hence the resonance pole arising after this merging can have a relatively small width. Note that this mechanism of producing the quasiparticle resonance is non-perturbative with respect to influence of the pairing correlation as is seen in the complex trajectories as a function of the pairing gap $\bar{\Delta}$. It is also different from the quasiparticle resonance associated with a deep-hole single-particle orbit (the hole-like quasiparticle resonance~\cite{Dobaczewski1996,Belyaev1987,Bulgac1980}), for which the pairing effect on the resonance width is known to be perturbative, i.e. the resonance width can be evaluated on the basis of the Fermi golden rule.

ii) \underline{Presence of four poles at close positions near the threshold:}  As seen in Fig.~\ref{fig4}, all the poles are located at positions with small absolute values in quasiparticle energy $|E|\lesssim 1$ MeV and in energy relative to the threshold $|e|=|E-|\lambda||\lesssim 1$ MeV. This situation is different from a typical resonance of the Breit-Wigner type, which is caused by an isolated single resonance pole. We shall examine influence of the four poles in next subsection.

iii) \underline{Four poles as two pairs:}  The trajectories shown in  Figs.~\ref{fig4} and \ref{fig5} suggest that the poles $a$ and $b$ may be regarded as a pair since, as the pairing strength increases, the poles $a$ and $b$ merge in the $E^{(2)}$-sheet and then they form a pair of resonance and anti-resonance. The other pair $\bar{a}$ and $\bar{b}$ is a qusiparticle-quasihole conjugate of the pair of $a$ and $b$ whose quasiparticle energies are negative ($\mathrm{Re}\left(E\right) <0$) in most cases. The pole pair $\bar{a}$ and $\bar{b}$ merges in the $E^{(3)}$-sheet. To distinguish the two pairs, we shall call the pair $a$ and $b$ the {\it quasiparticle poles} while $\bar{a}$ and $\bar{b}$ the {\it quasihole poles}. The roles of these two pairs of poles will be discussed just below.

We shall analyze here how the presence of four poles close to the threshold affect the physical observable, the elastic cross section. For this purpose, we use approximated expressions of the S-matrix in which contributions of the poles are explicitly introduced, but in a simplified way. We term it the pole approximation of the S-matrix. If we consider only one pole, then the approximated S-matrix is given as~\cite{Taylor,Kukulin,Sitenko}
\begin{equation}
S^{\mathrm{pole}}_{s_{1/2}}=e^{2i\delta^{\mathrm{bg}}}\frac{k-k_{\mathrm{pole}}^{\ast}}{k-k_{\mathrm{pole}}}.
\label{smatrix_approx}
\end{equation}
where $k$ is the wave number of scattering particle and $k_{pole}$ is the wave number of a pole. In the present analysis, we introduce an approximate S-matrix, expressed in terms of the asymptotic wave number $k_1$ of the upper (particle) component $\varphi_{1}(r)$ of the quasiparticle wave function. In the case when we consider all the {\it four} poles ($a$$\bar{a}$$b$$\bar{b}$), we generalize Eq.~(\ref{smatrix_approx}) and we assume $S^{\mathrm{pole}}_{s_{1/2}}$ being a product of individual contributions of poles: \footnote{With this simple approximation we intend to have qualitative understanding of the S-matrix poles rather than to reproduce the exact numerical result. For more precise and quantitative analysis of contributions of the poles, it may be necessary to treat explicitly the two-channel character of the problem, and introduce parametrizations suitable for multi-channel S-matrix (e.g. Refs.~\cite{Taylor, Kukulin}). We leave it for future study.}
\begin{equation}
S^{\mathrm{pole}}_{s_{1/2}}=\frac{k_1-\bar{k}^{\ast}_{1a}}{k-\bar{k}_{1a}}\cdot\frac{k_1-\bar{k}^{\ast}_{1b}}{k_1-\bar{k}_{1b}}\cdot\frac{k_1-\bar{k}^{\ast}_{1\bar{a}}}{k_1-\bar{k}_{1\bar{a}}}\cdot\frac{k_1-\bar{k}^{\ast}_{1\bar{b}}}{k_1-\bar{k}_{1\bar{b}}}.
\label{smatrix_approx4}
\end{equation}
The elastic cross section and the phase shift are expressed with $S^{\mathrm{pole}}_{s_{1/2}}$ as below:
\begin{equation}
\sigma^{\mathrm{pole}}_{s_{1/2}}=\frac{\pi}{k_1^{2}}\left|S^{\mathrm{pole}}_{s_{1/2}}-1\right|^2 ,
\end{equation}
\begin{equation}
\delta^{\mathrm{pole}}_{s_{1/2}}=\frac{1}{2}\arccos\left( \mathrm{Re}\left( S^{\mathrm{pole}}_{s_{1/2}}\right) \right).
\label{phaseshift_approx}
\end{equation}
When we consider two poles, we use a product form similar to Eq.~(\ref{smatrix_approx4}) but including only two terms corresponding to the relevant poles. 

Let us first discuss the case of $\bar{\Delta}=1.5$ MeV where the positions of the four poles are off the real and the imaginary axes, and then the case of $\bar{\Delta}=1.0$ MeV, where all the pole positions are on the imaginary axis.

\begin{figure}[h]
\begin{minipage}{0.48\hsize}
\includegraphics[width=77mm]{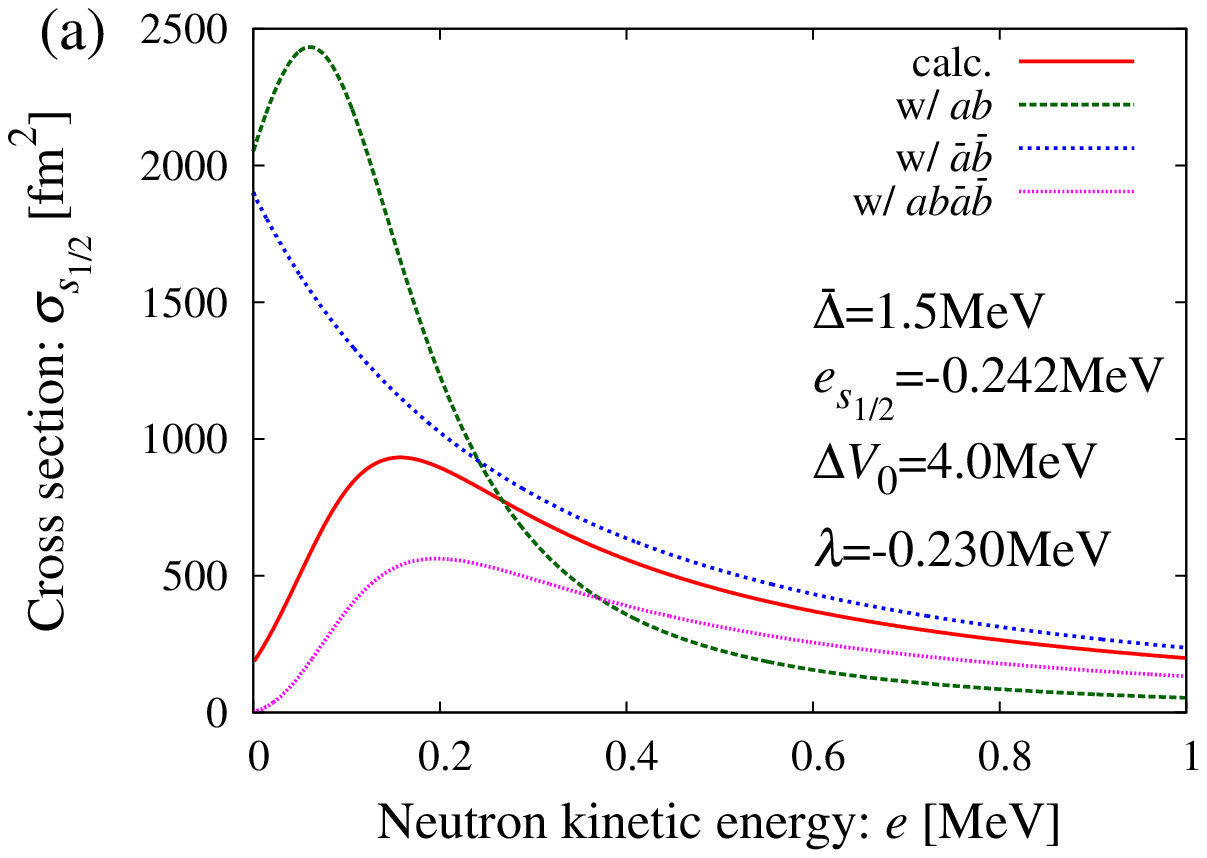}
\end{minipage}
\begin{minipage}{0.48\hsize}
\includegraphics[width=77mm]{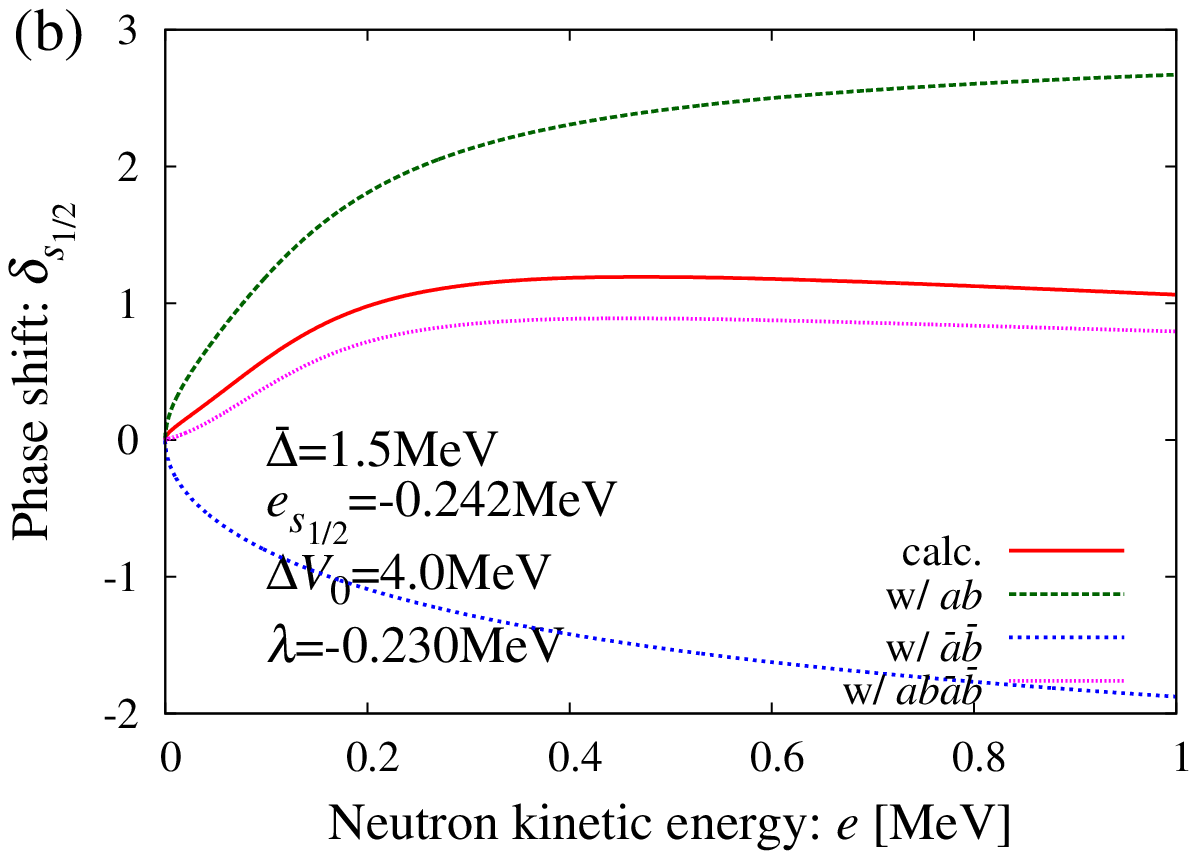}
\end{minipage}
\caption{Influence of the S-matrix poles on the elastic cross section $\sigma_{s_{1/2}}$ and the phase shift $\delta_{s_{1/2}}$, evaluated using the pole approximation (\ref{smatrix_approx4}) which includes the four poles $a$, $b$, $\bar{a}$ and $\bar{b}$ (magenta dotted curve),  the pole pair $a$ and $b$ (green dotted curve), or the other pole pair $\bar{a}$ and $\bar{b}$ (blue dotted curve). The results obtained with direct numerical calculation are shown with red solid curve. The potential parameter and the Fermi energy are $\Delta V_0=4.0$ MeV ($e_{2s_{1/2}}=-0.242$ MeV) and $\lambda=-0.230$ MeV with the average pairing gap  $\bar{\Delta}=1.5$ MeV.}
\label{fig7}
\end{figure}

Figure~\ref{fig7} shows the elastic cross section $\sigma^{\mathrm{pole}}_{s_{1/2}}$ and the phase shift $\delta^{\mathrm{pole}}_{s_{1/2}}$ obtained with the pole approximation of the S-matrix (plotted as a function of the neutron kinetic energy $e=E-|\lambda|=\hbar^2k_1^2/2m$). We show results of three different choices of poles; i) the case where only the resonance pole $a$ and the anti-resonance $b$ are selected, ii) the case where the remaining two poles $\bar{a}$ and $\bar{b}$ are selected, and iii) all the four poles $a$, $b$, $\bar{a}$, and $\bar{b}$ are included. It is seen that the approximation including the poles $a$ and $b$ gives a peak in the cross section at $e\approx 0.061$ MeV. The resonance structure is caused by the pole $a$ (and $b$), i.e., the peak energy corresponds well to the pole positions $e_{\mathrm{res}}=\mathrm{Re}\left(E_{a,b}\right)-|\lambda|=0.0606$ ($|\lambda|=0.230$ MeV) of $a$ and $b$. We emphasize also that contribution from another pair, $\bar{a}$ and $\bar{b}$, is important as well to better reproduce the exact numerical results, although it causes no resonance behaviors. It is noticed that the phase shift caused by the pair $\bar{a}$ and $\bar{b}$ is not small, varying  sizably  from 0 to $\sim -\pi/2$ in the considered energy range $e=0 \sim 0.5$ MeV, This is because the pole positions of $\bar{a}$ and $\bar{b}$ are close to the threshold $E=|\lambda|, e=0$ (The distance in the $E$-plane is as small as $|E_{\bar{a},\bar{b}}-|\lambda||=0.539$ MeV). Consequently, as a combined effect of the two pairs of poles (the four poles), the peak energy of the cross section is shifted to a slightly higher energy $e=0.196$ MeV in the pole approximation (magenta curve in Fig.~\ref{fig7} (a)) and $e=0.158$ MeV in the exact numerical result (red curve), while the resonance structure remains.

\begin{figure}[h]
\begin{minipage}{0.48\hsize}
\includegraphics[width=77mm]{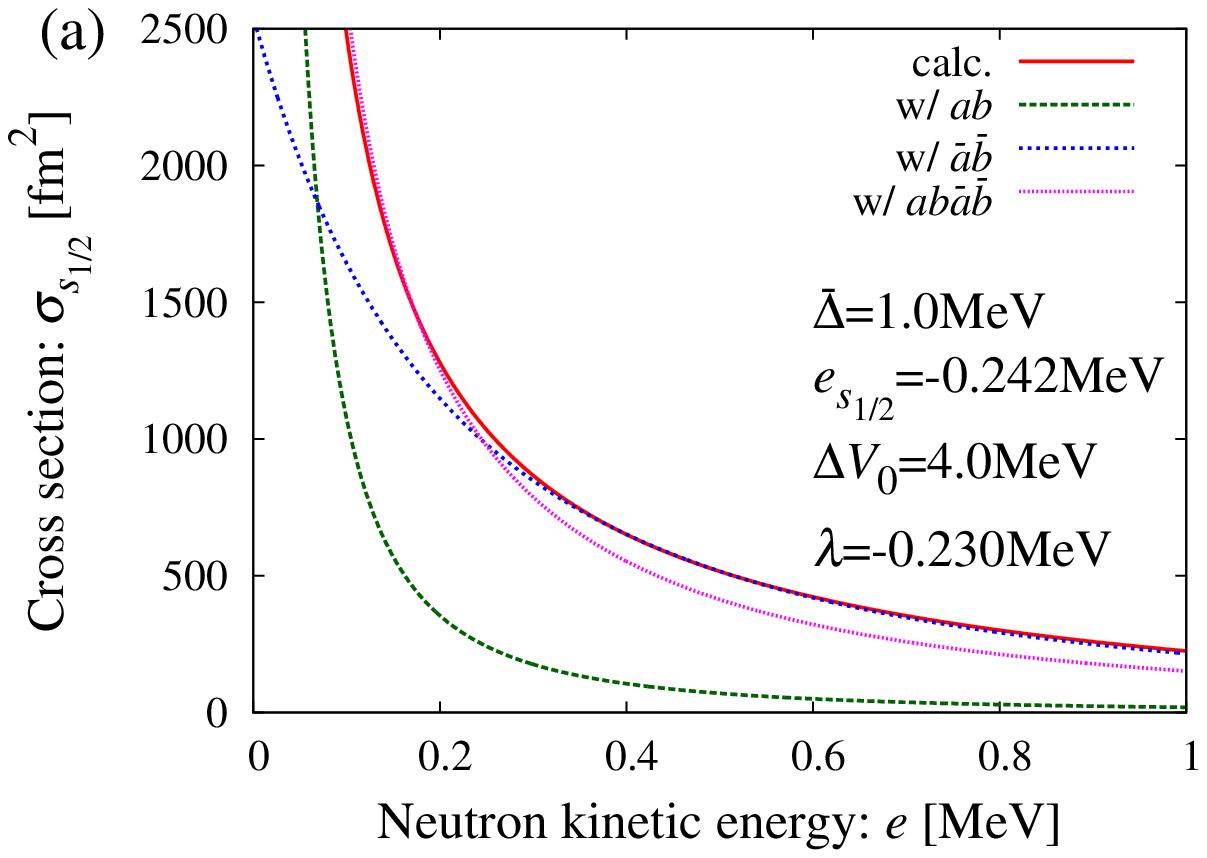}
\end{minipage}
\begin{minipage}{0.48\hsize}
\includegraphics[width=77mm]{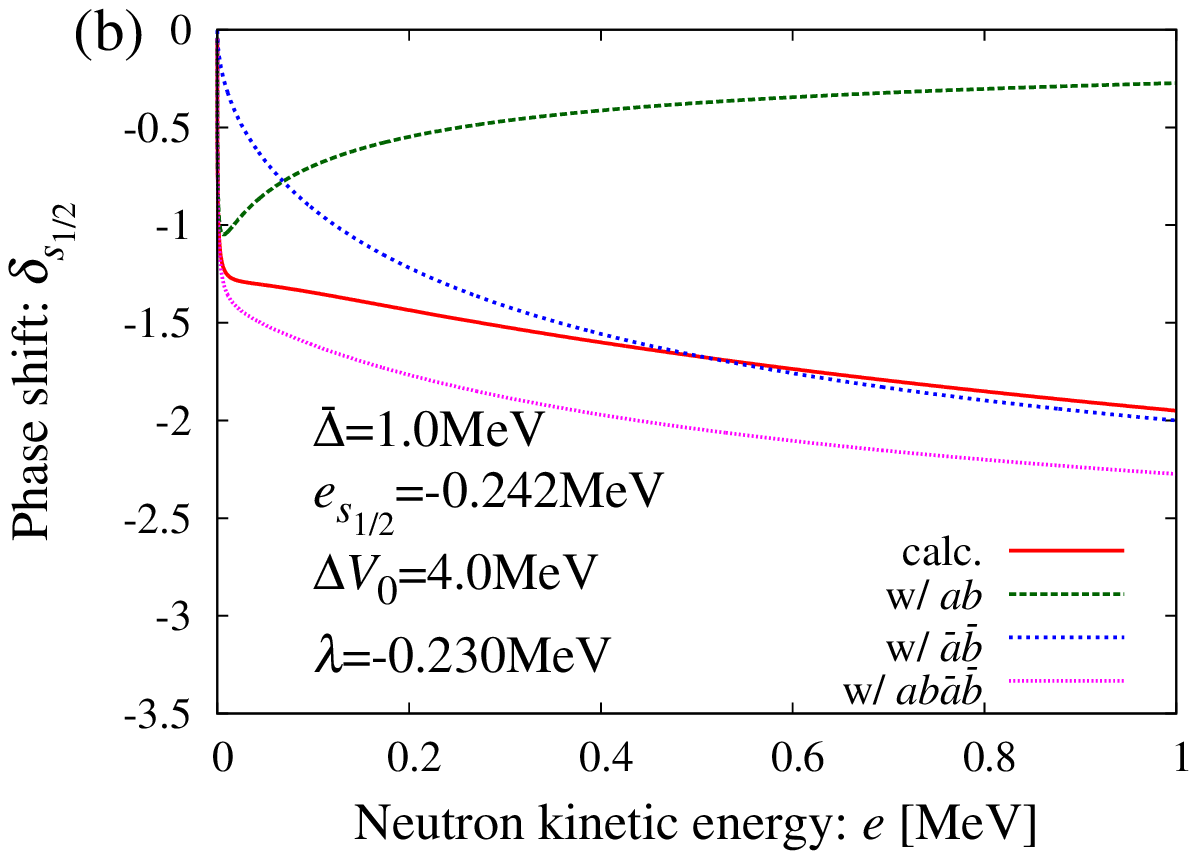}
\end{minipage}
\caption{The same as Fig.~\ref{fig7}, but for the average pairing gap $\bar{\Delta}=1.0$ MeV.}
\label{fig8}
\end{figure}

Figure~\ref{fig8} shows the analysis of the case of $\bar{\Delta}=1.0$ MeV. In this case, all the four poles are located on the imaginary $k_1$ and $k_2$ axes, and the positions in the $E$-plane are $E_{a}=0.2295$ and $E_{\bar{a}}=-0.2295$ MeV on the first Riemann sheet, $E_{b}=0.1376$ MeV on the second sheet, and $E_{\bar{b}}=-0.1376$ MeV on the third sheet. 

It is seen in the green dotted curve in Fig.~\ref{fig8} that the contribution of the pole pair $a$ and $b$ causes a divergent behavior at $e=0$ in the elastic cross section, and it is mainly due to the pole $a$, which plays a role of a virtual state. The characteristic feature of virtual state is also seen in the contribution to the phase shift, which drops steeply around $e=0$. The pole $b$ also contributes as an uprise of the phase shift after the steep drop. The contribution of the other pole pair, $\bar{a}$ and $\bar{b}$, shown with blue dotted curve, is similar to that in the case of $\bar{\Delta}=1.5$ MeV. We notice also that the features of the virtual state remains in the evaluation including all the four poles, and that the numerically obtained cross section and the phase shift (red curves) can be understood as a combined effect of the contributions from the two pairs of the poles.

The different contributions from the two pole pairs can be stated with more physical terms. The Bogoliubov quasiparticle states appear both as negative energy eigenstates as well as positive energy ones, and the former is the quasihole states. In the present description, the pole pair $a$ and $b$ stems from the positive energy quasiparticle state associated with the weakly bound $2s_{1/2}$ single-particle state.  The other pole pair $\bar{a}$ and $\bar{b}$ stems from the conjugate negative energy quasihole state. The above analysis indicates that both the positive energy quasiparticle state (the quasiparticle poles $a$ and $b$) and the negative energy quasihole state (the quasihole poles $\bar{a}$ and $\bar{b}$) contribute to the neutron elastic scattering. This arises from the peculiar feature of the Bogoliubov quasiparticle, which has a double character as a particle and as a hole, which are  coupled to one another by exchanging a Cooper pair with the pair potential. It is noted that the positive energy quasiparticle state ($E>0$) can pass through the threshold since the corresponding kinetic energy of neutron $e=\lambda+E$ can become both negative (bound) and positive (unbound) depending on the pairing gap. Consequently, the quasparticle poles $a$ (and $b$) can become either a weakly bound state, a virtual state or a resonance (anti-resonance). On the other hand the  energy $e=\lambda-E <  \lambda < 0$ of the quasihole state relative to the threshold is always negative, i.e. the quasihole states are bound to the nuclear potential. In the case of near-drip-line nuclei (with shallow Fermi energy or small $|\lambda|$) the binding energy of the quasihole state is small, and hence the contribution of the quasihole state (the quasihole poles $\bar{a}$ and $\bar{b}$) becomes sizable as is seen in the above analysis.

\subsection{Systematics of resonance poles}
The scattering property and the resonance structure depend not only on the average pairing gap $\bar{\Delta}$ but also on the single-particle energy $e_{2s_{1/2}}$ (controlled by the potential depth $\Delta V_0$), and the Fermi energy $\lambda$. We shall discuss systematical behavior against the change in both $\bar{\Delta}$ and $\Delta V_0$, focusing on the resonance pole. We continue discussing the extreme case of the very shallow Fermi energy $\lambda=-0.230$ MeV as above. We treat the case of different choice of the Fermi energy in the next subsection.

\begin{figure}[h]
\begin{center}
\includegraphics[width=90mm,angle=0]{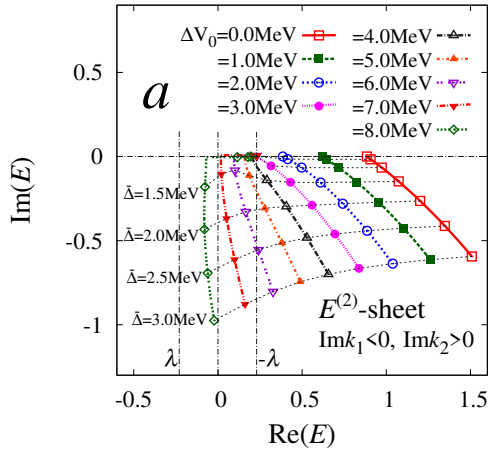}
\end{center}
\caption{Systematics of trajectories of the pole $a$ on the $E^{(2)}$-sheet. The average pairing gap $\bar{\Delta}$ is changed continuously in the range $\bar{\Delta}=0 - 3$ MeV for various values of the potential depth $\Delta V_0=0.0,~ 1.0,~ 2.0, \cdots 8.0$ MeV. The Fermi energy is fixed to $\lambda=-0.230$ MeV.}
\label{fig9}
\end{figure}

Figure~\ref{fig9} shows the trajectories of the quasiparticle resonance pole $a$ emerging on the $E^{(2)}$-sheet with systematic variation of $0< \bar{\Delta} < 3.0$ MeV and $0<\Delta V_{0}<8.0$ MeV. We classify the results in three cases where 1) the single-particle energy $e_{2s_{1/2}}$ is negative and satisfies $e_{2s_{1/2}}<2\lambda$, 2) $e_{2s_{1/2}}$ is negative and satisfies $2\lambda < e_{2s_{1/2}}<0$, and 3) the $2s_{1/2}$ orbit is unbound (a virtual state).

\noindent
\underline{1) The case of $e_{2s_{1/2}}<2\lambda=-0.460$ MeV}

This corresponds to $\Delta V_{0}=0.0, 1.0, 2.0$ MeV, and this is a situation where binding of the $2s_{1/2}$ obit is relatively strong. As a representative, we shall examine the case of $e_{2s_{1/2}}=-1.115$ MeV ($\Delta V_0=0$ MeV).

\begin{figure}[h]
\begin{center}
\begin{minipage}{0.48\hsize}
\includegraphics[width=90mm]{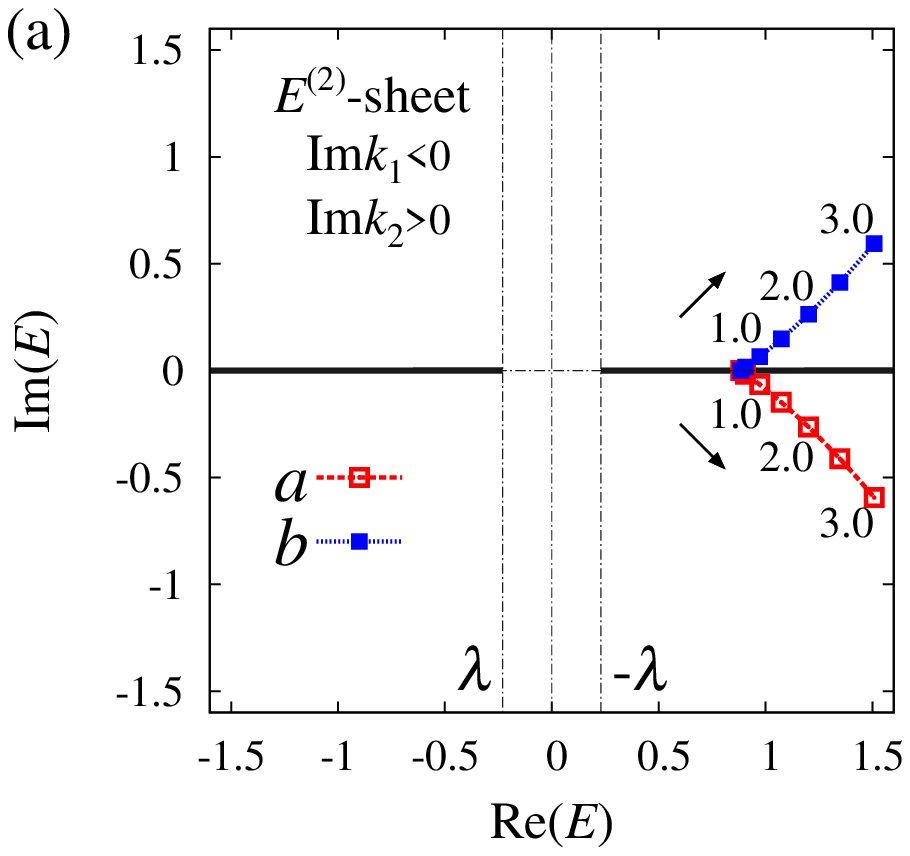}
\end{minipage}
\begin{minipage}{0.48\hsize}
\includegraphics[width=90mm]{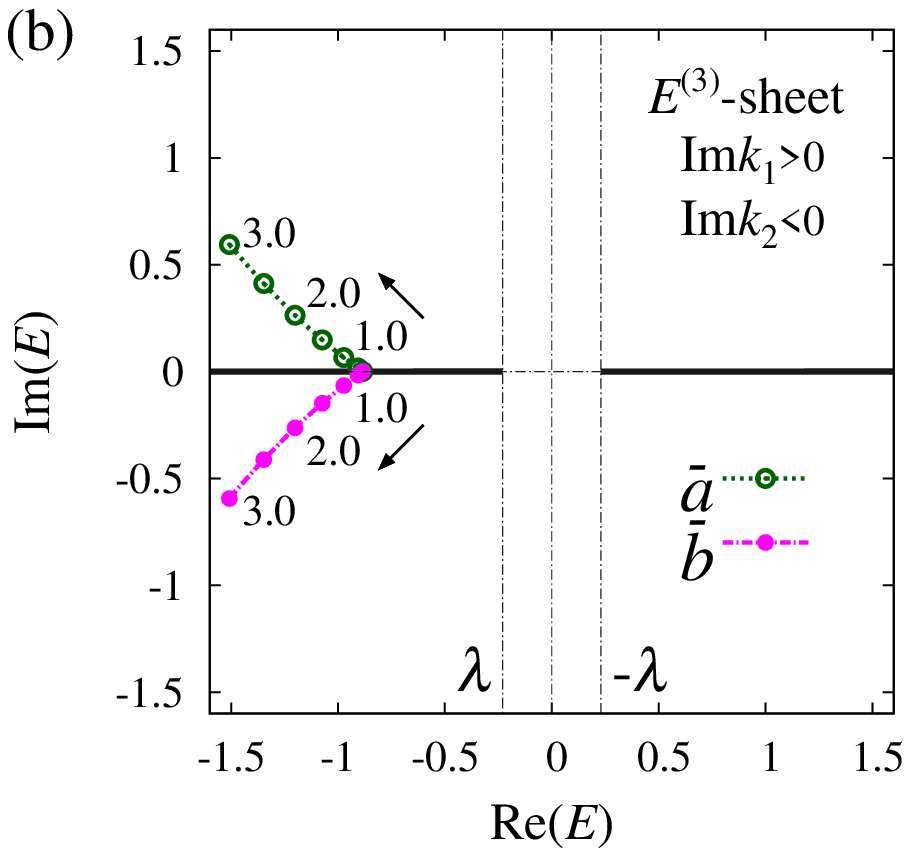}
\end{minipage}
\begin{minipage}{0.48\hsize}
\includegraphics[width=90mm]{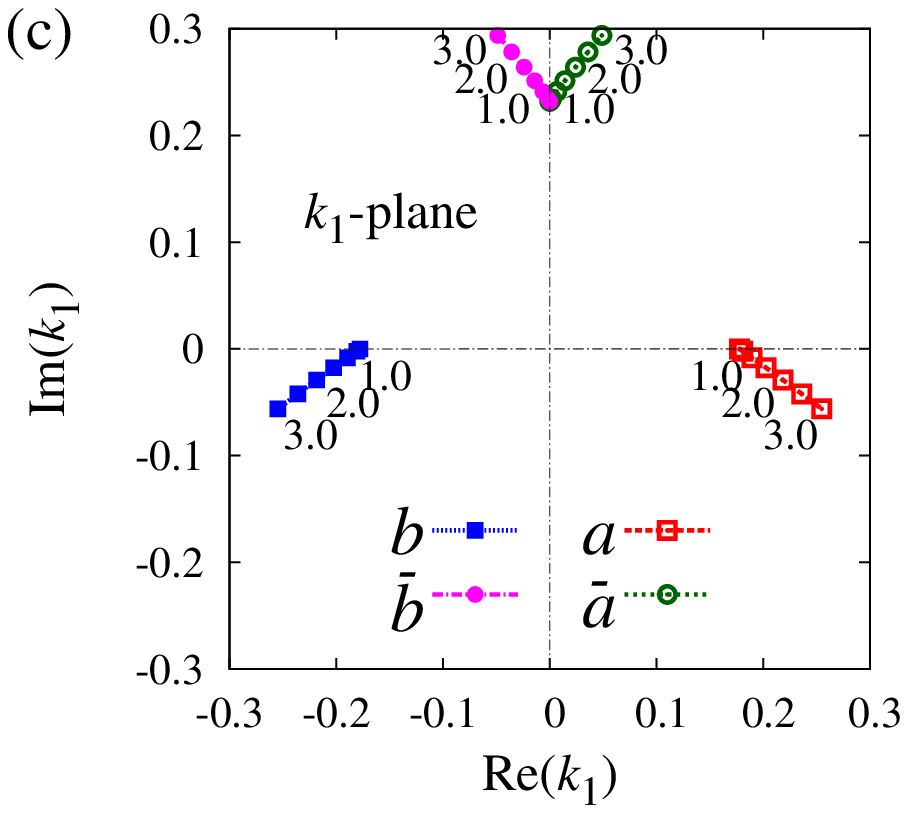}
\end{minipage}
\end{center}
\caption{Pole trajectories in the complex $E$- and $k_{1}$-planes, where the average pairing gap $\bar{\Delta}$ is varied in the range $\bar{\Delta}=0 - 3$ MeV while the $2s_{1/2}$ single-particle energy and the Fermi energy are fixed to $e_{2s_{1/2}}=-1.115$ MeV ($\Delta V_0=0.0$ MeV) and $\lambda=-0.230$ MeV.}
\label{fig10}
\end{figure}

Figure~\ref{fig10} shows the trajectory of the four poles $a,b,\bar{a}$ and $\bar{b}$ as a function of  $\bar{\Delta}$ on the $E$- and $k_1$-planes. For $e_{2s_{1/2}}<2\lambda$ the positive energy quasiparticle state in the zero pairing limit $\bar{\Delta}\rightarrow 0$ appears on the positive real $E$-axis above the threshold energy $E>-\lambda$, but in the unphysical $E^{(2)}$-sheet. It represents a hole state embedded in the continuum.  With increasing $\bar{\Delta}$, the pole immediately leaves the real $E$-axis and moves on the $E^{(2)}$-sheet with gradual increase of the imaginary part $\mathrm{Im}\left(E_a\right)$. This corresponds to the case known as the hole-like quasiparticle resonance. From the position of the pole $a$, we find a small width of the quasiparticle resonance $\Gamma=2\mathrm{Im}(E)=0.0-1.2$ MeV for $\bar{\Delta}=0.0-3.0$ MeV. Dependence of the width on $\bar{\Delta}$ is approximately quadratic, and this dependence is consistent with a perturbative estimate of the hole-like quasiparticle resonance~\cite{Dobaczewski1996,Belyaev1987,Bulgac1980} based on the Fermi golden rule.  

\begin{figure}[h]
\begin{minipage}{0.48\hsize}
\includegraphics[width=77mm]{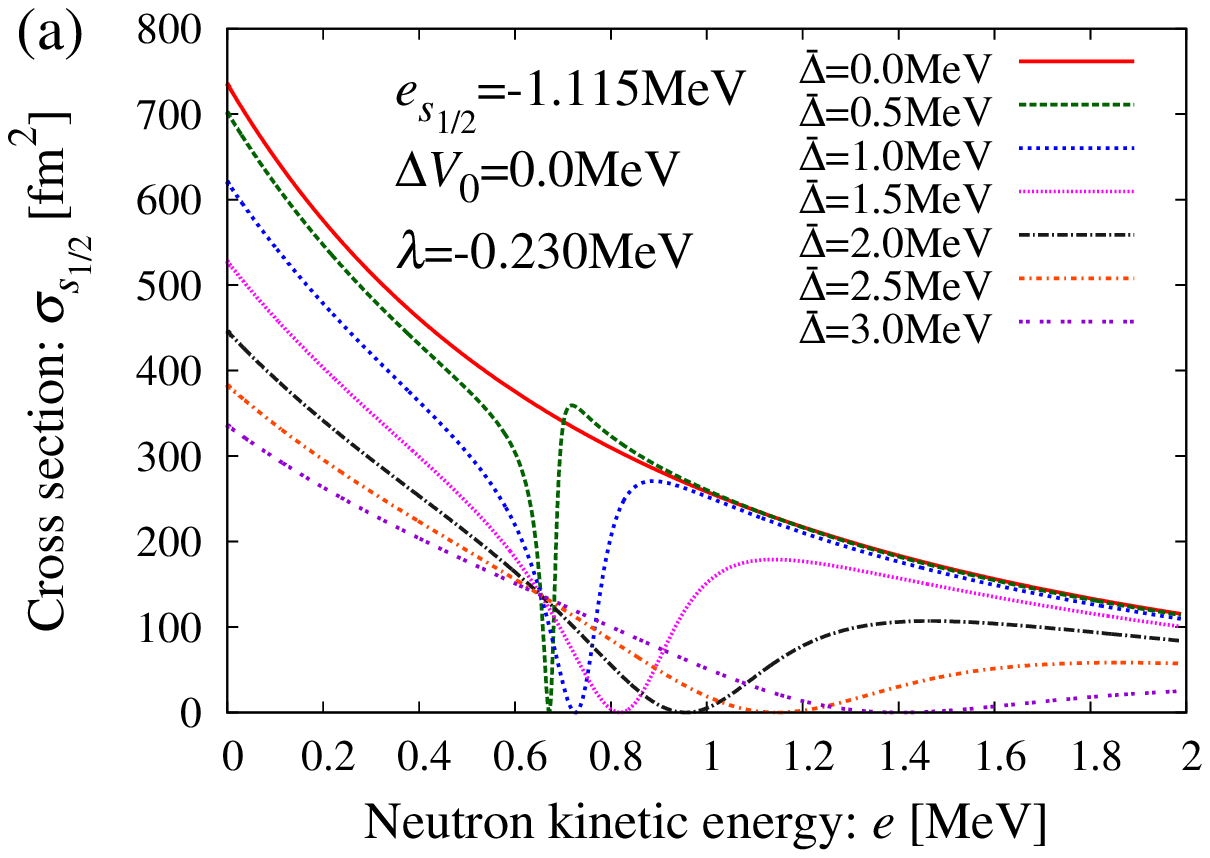}
\end{minipage}
\begin{minipage}{0.48\hsize}
\includegraphics[width=77mm]{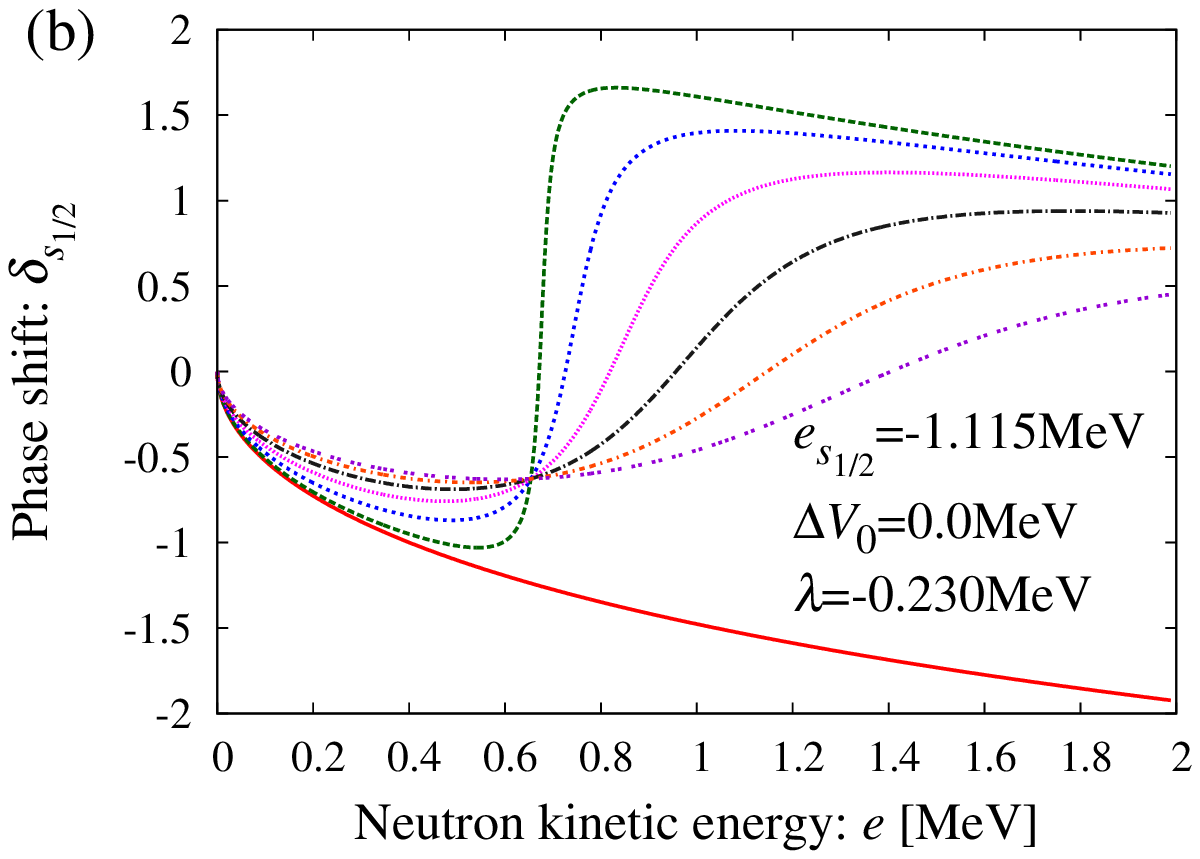}
\end{minipage}
\caption{The same as Fig.~\ref{fig1}, but for $e_{2s_{1/2}}=-1.115$ MeV ($\Delta V_0=0.0$ MeV).}
\label{fig11}
\end{figure}

\begin{figure}[h]
\begin{minipage}{0.48\hsize}
\includegraphics[width=77mm]{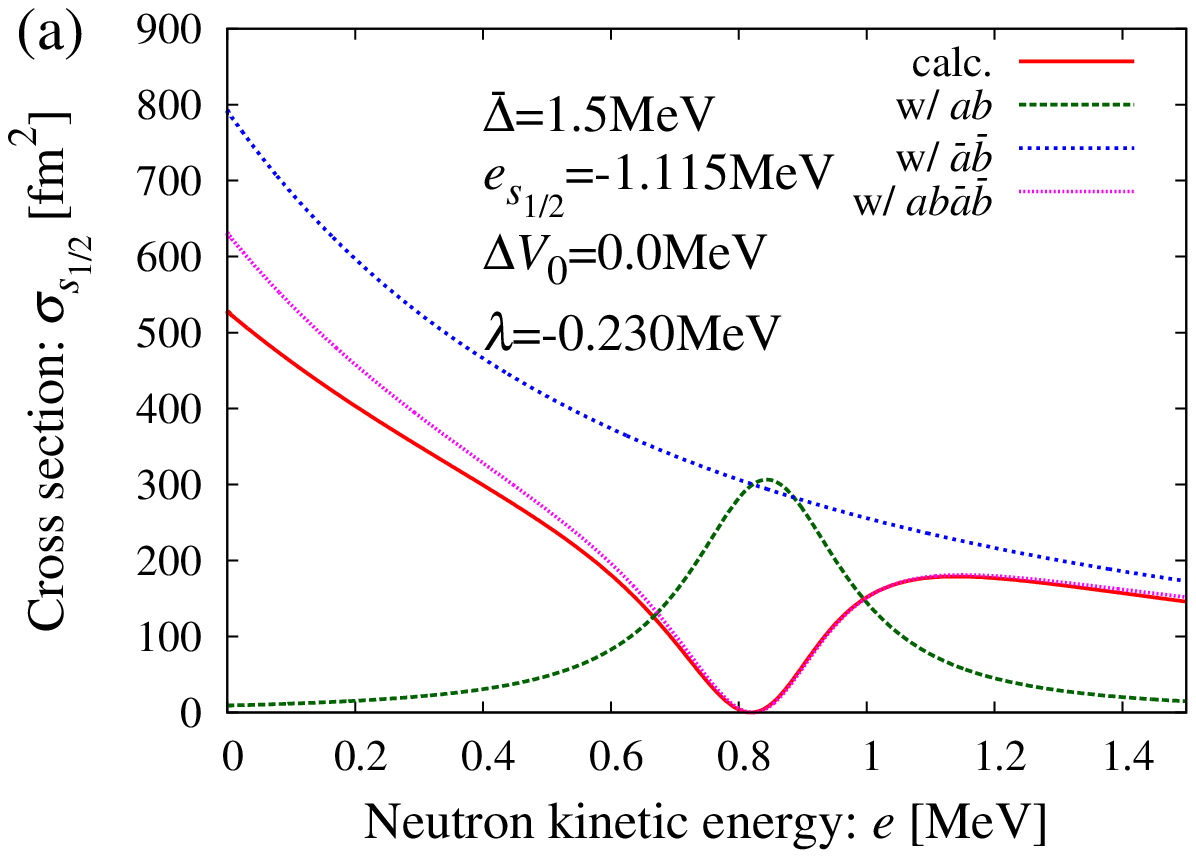}
\end{minipage}
\begin{minipage}{0.48\hsize}
\includegraphics[width=77mm]{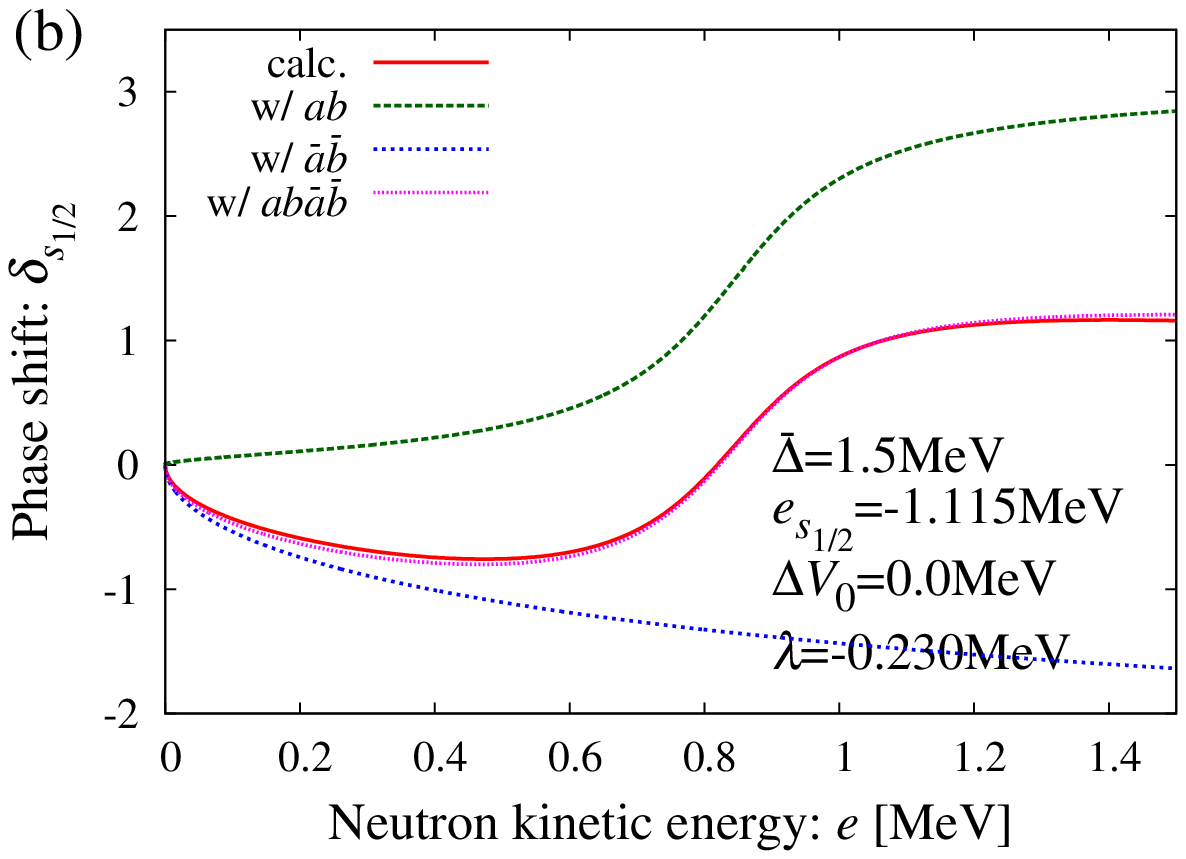}
\end{minipage}
\caption{The same as Fig.~\ref{fig7}, but for $e_{2s_{1/2}}=-1.115$ MeV ($\Delta V_0=0.0$ MeV) and $\bar{\Delta}=1.5$ MeV.}
\label{fig12}
\end{figure}

The elastic cross section and the phase shift is shown in Fig.~\ref{fig11} for various values of $\bar{\Delta}=0.0-3.0$ MeV. The quasiparticle resonance is seen clearly in the phase shift. However the resonance appears in the cross section not as a peak but as a minimum, and it is because of the presence of a ``background'' phase shift. The origin of the background phase shift is the quasihole poles $\bar{a}$ and $\bar{b}$, as is demonstrated in Fig.~\ref{fig12}, where the contributions of the pole pair $a$ and $b$ and the other pair $\bar{a}$ and $\bar{b}$ are separately shown using the pole approximation (cf. Eq.~(\ref{smatrix_approx4})). It is seen that the background phase shift caused by the quasihole poles is sizable. Consequently the combination of the resonance and the background produces the minimum of the cross section around the resonance energy.

\noindent
\underline{2) The case of $2\lambda<e_{2s_{1/2}}<0$}

This case corresponds to $\Delta V_{0}=3.0, 4.0, 5.0, 6.0$ MeV. A typical example of $\Delta V_{0}=4.0$ MeV is discussed in detail in subsection \ref{subsec:trajectory} in connection with Figs.~\ref{fig4} and \ref{fig5}. In all these cases, we find the non-perturbative variation of the pole $a$ as a function of the pairing strength $\bar{\Delta}$ with transition from a weakly bound state $\rightarrow$ a virtual state $\rightarrow$ a resonance pole.

Dependence of the pole $a$ on $\Delta V_0$ (hence the $2s_{1/2}$ single-particle energy $e_{2s_{1/2}}$) is seen in Fig.~\ref{fig9}. With increasing $\Delta V_0$ from 3.0 to 6.0 MeV, the single-particle energy $e_{2s_{1/2}}$ varies as $e_{2s_{1/2}}=-0.411, -0.242, -0.113, -0.029$ MeV. In the case of $\Delta V_0=3.0$ MeV (where $e_{2s_{1/2}}=-0.411$ MeV is close to $2\lambda=-0.460$ MeV), the pole $a$ leaves the real $E$-axis and the imaginary $k_1$-axis at a position which is very close to the threshold $E=-\lambda$ ($k_1=0$). Hence the pole $a$ gives rise a narrow resonance in this case. With increasing $\Delta V_0$ ($e_{2s_{1/2}}$), the trajectory of the pole $a$ moves leftward (i.e. it is located more distant from the threshold). For $\Delta V_0=4.0$ MeV ($e_{2s_{1/2}}=-0.242$ MeV), the pole $a$ becomes a resonance with moderate width as we already discussed. Note that the $2s_{1/2}$ single-particle energy is lower than the Fermi energy, $e_{2s_{1/2}} \lesssim \lambda$, in these cases. For $\Delta V_0=5.0$ and 6.0 MeV, where $e_{2s_{1/2}} \gtrsim \lambda$, the trajectory of the resonance pole $a$ is more distant from the threshold, and hence we find that the pole $a$ does not show up as a visible resonance in observables. The resonance with a narrow or moderate width emerges under the condition $e_{2s_{1/2}} \lesssim\lambda$.

\noindent
\underline{3) The case where the $2s_{1/2}$ orbit is unbound}

This corresponds to $\Delta V_{0}=7.0, 8.0$ MeV in Fig.~\ref{fig9}. In this situation, the $2s_{1/2}$ orbit in the zero pairing limit is a virtual state, appearing in the unphysical $E^{(2)}$-sheet.

\begin{figure}[h]
\begin{minipage}{0.48\hsize}
\includegraphics[width=90mm]{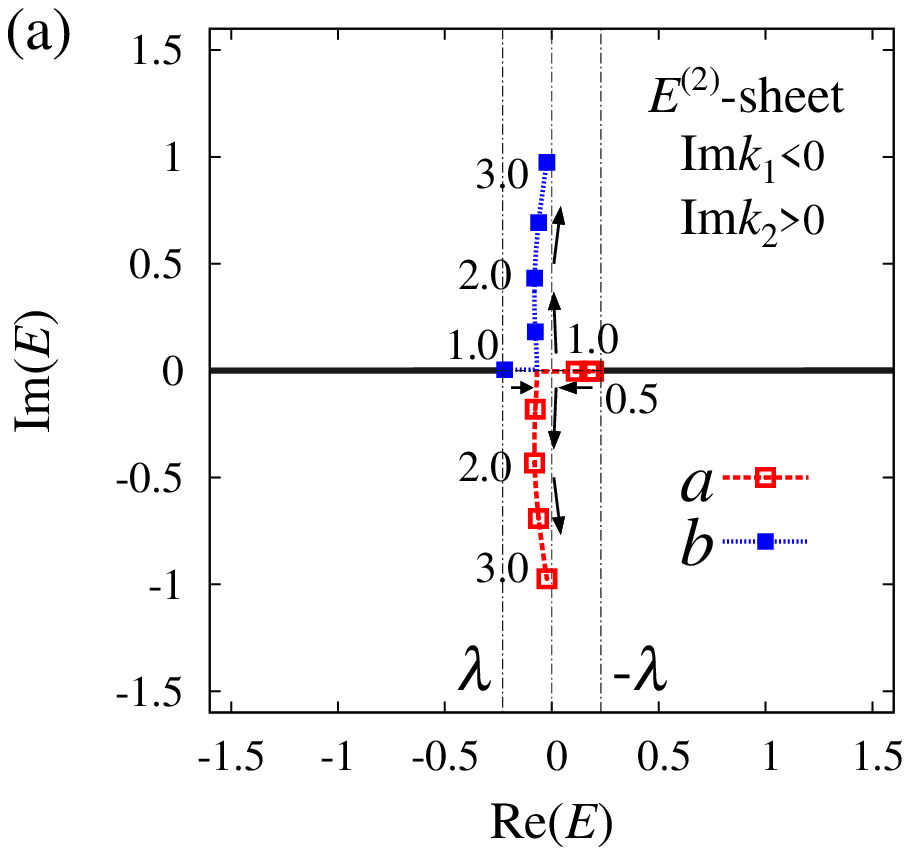}
\end{minipage}
\begin{minipage}{0.50\hsize}
\includegraphics[width=90mm]{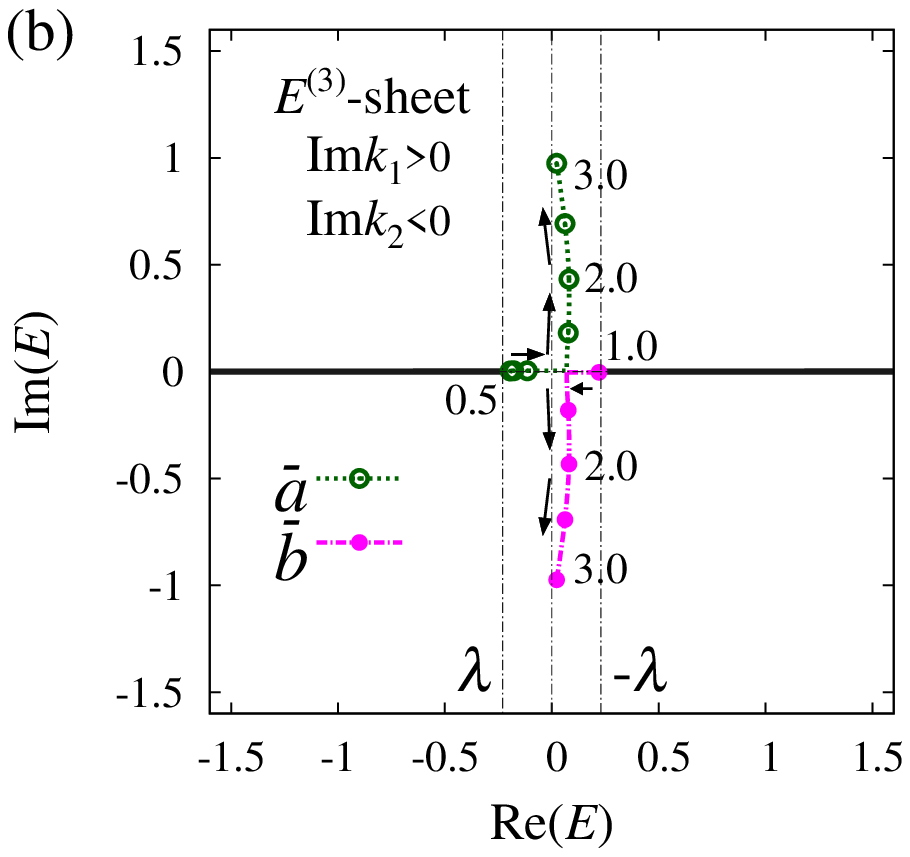}
\end{minipage}
\begin{minipage}{0.48\hsize}
\includegraphics[width=90mm]{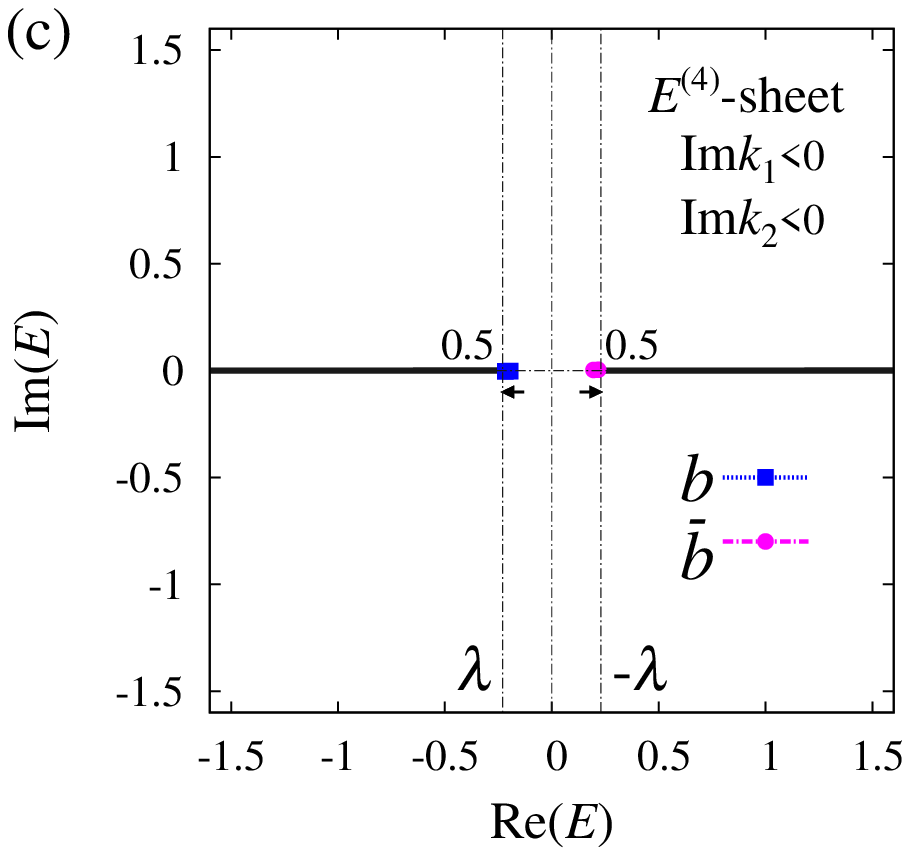}
\end{minipage}
\begin{minipage}{0.48\hsize}
\includegraphics[width=90mm]{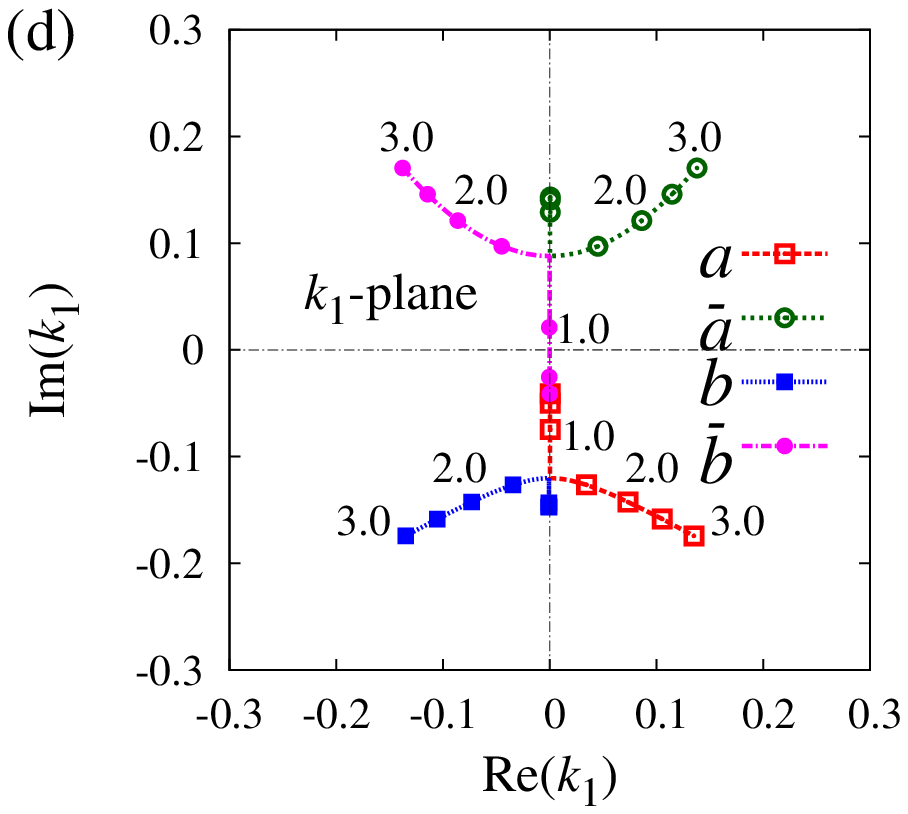}
\end{minipage}
\caption{Pole trajectories in the complex $E$- and $k_{1}$-planes, where the $2s_{1/2}$ single-particle state is unbound (being a virtual state) with $\Delta V_0=8.0$ MeV while the average pairing gap $\bar{\Delta}$ is varied in the range $\bar{\Delta}=0 - 3$ MeV. The Fermi energy are fixed to $\lambda=-0.230$ MeV.}
\label{fig13}
\end{figure}

Figure~\ref{fig13} shows how this virtual state is affected by the pairing correlation. As $\bar{\Delta}$ increases from 0 to 1.303 MeV, the pole $a$ remains as a virtual state, but moves further away from the threshold ($k_1=0$) along the negative imaginary $k_1$-axis. After merging with the pole $b$ (with $\bar{\Delta}=1.303$ MeV), it becomes a resonance pole in a similar way to the case 2). Note however that the resonance pole $a$ is more distant from the threshold than in the case 2).

\begin{figure}[h]
\begin{minipage}{0.48\hsize}
\includegraphics[width=77mm]{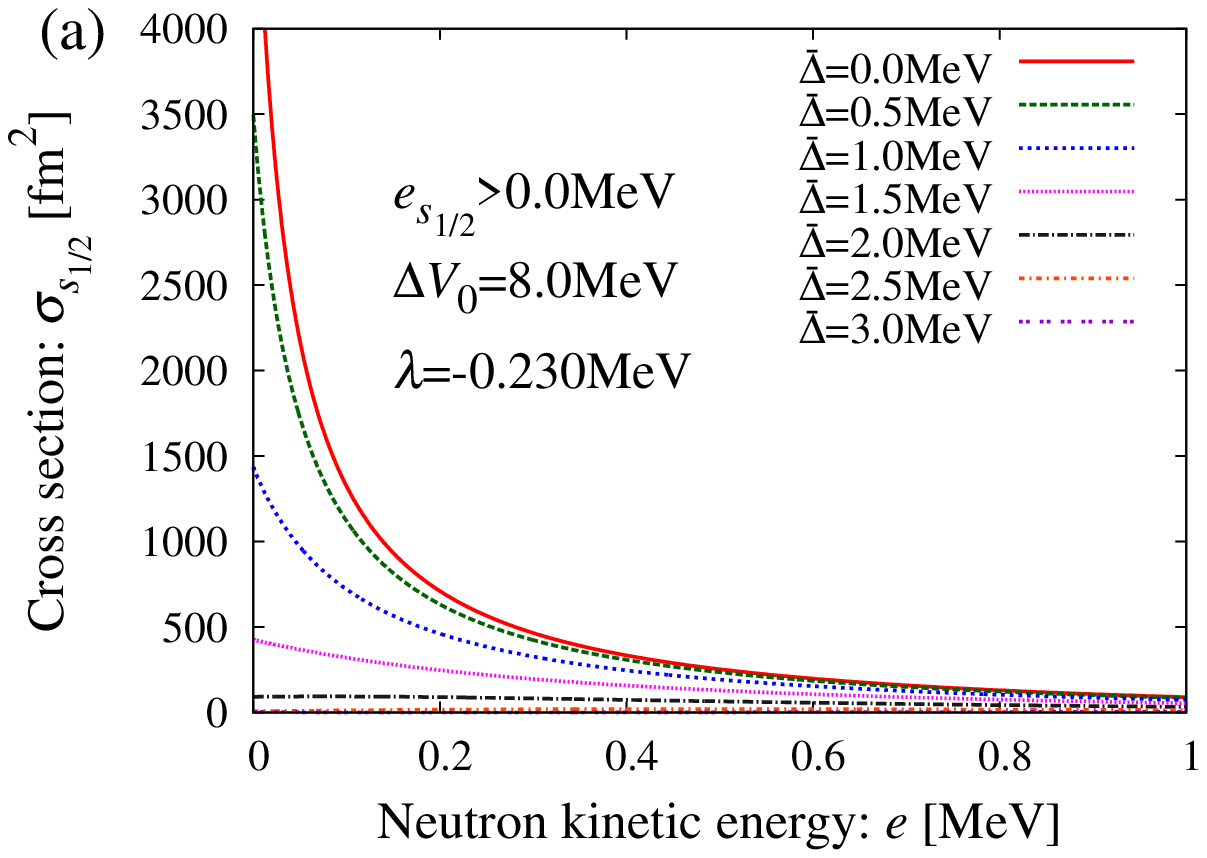}
\end{minipage}
\begin{minipage}{0.48\hsize}
\includegraphics[width=77mm]{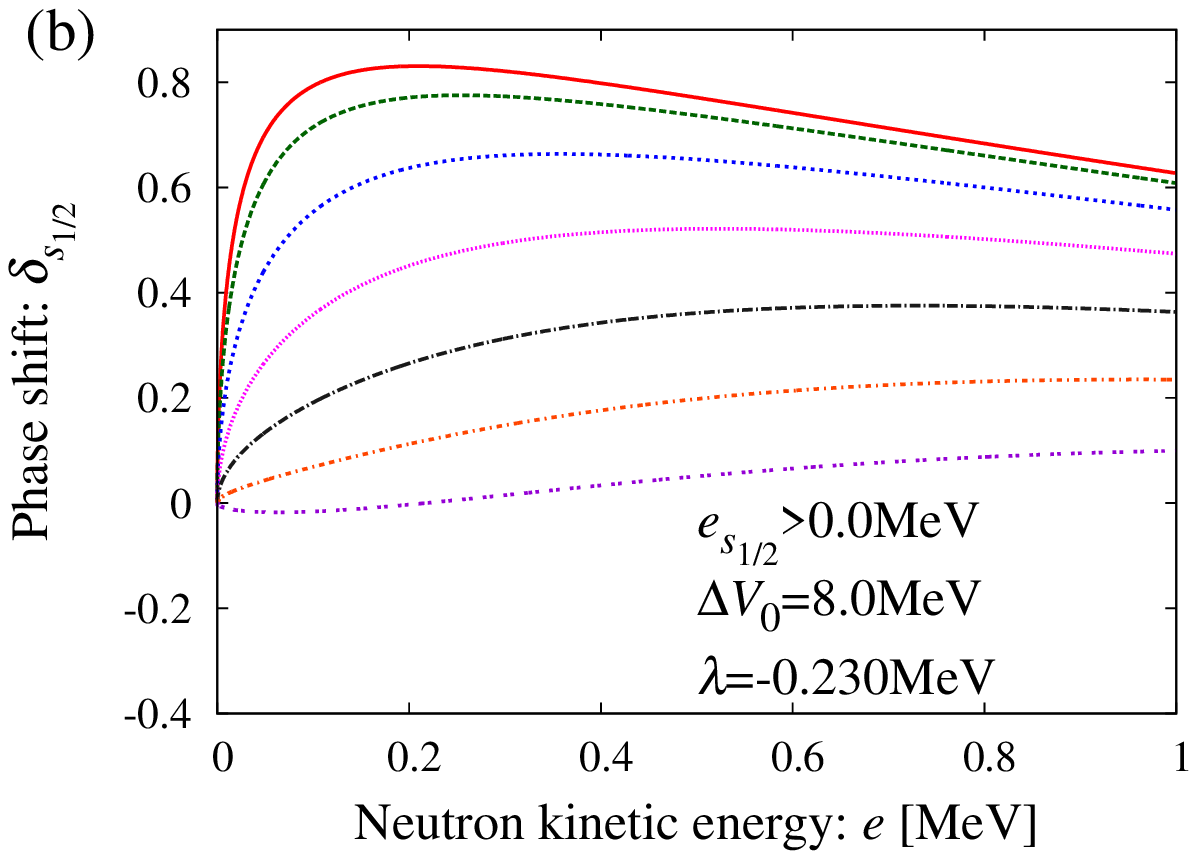}
\end{minipage}
\caption{The same as Fig.~\ref{fig1}, but for $\Delta V_0=8.0$ MeV, with which the $2s_{1/2}$ single-particle state is unbound (being a virtual state).}
\label{fig14}
\end{figure}

\begin{figure}[h]
\begin{minipage}{0.48\hsize}
\includegraphics[width=77mm]{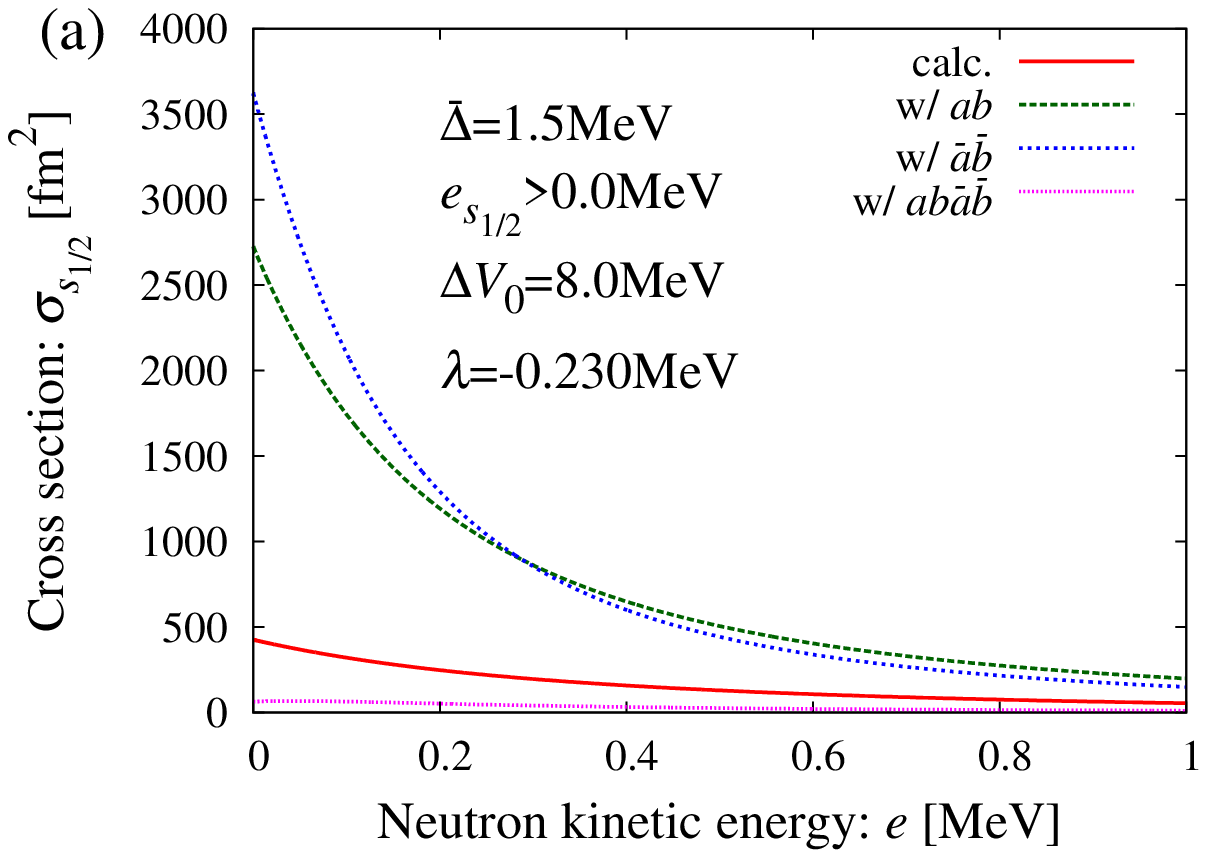}
\end{minipage}
\begin{minipage}{0.48\hsize}
\includegraphics[width=77mm]{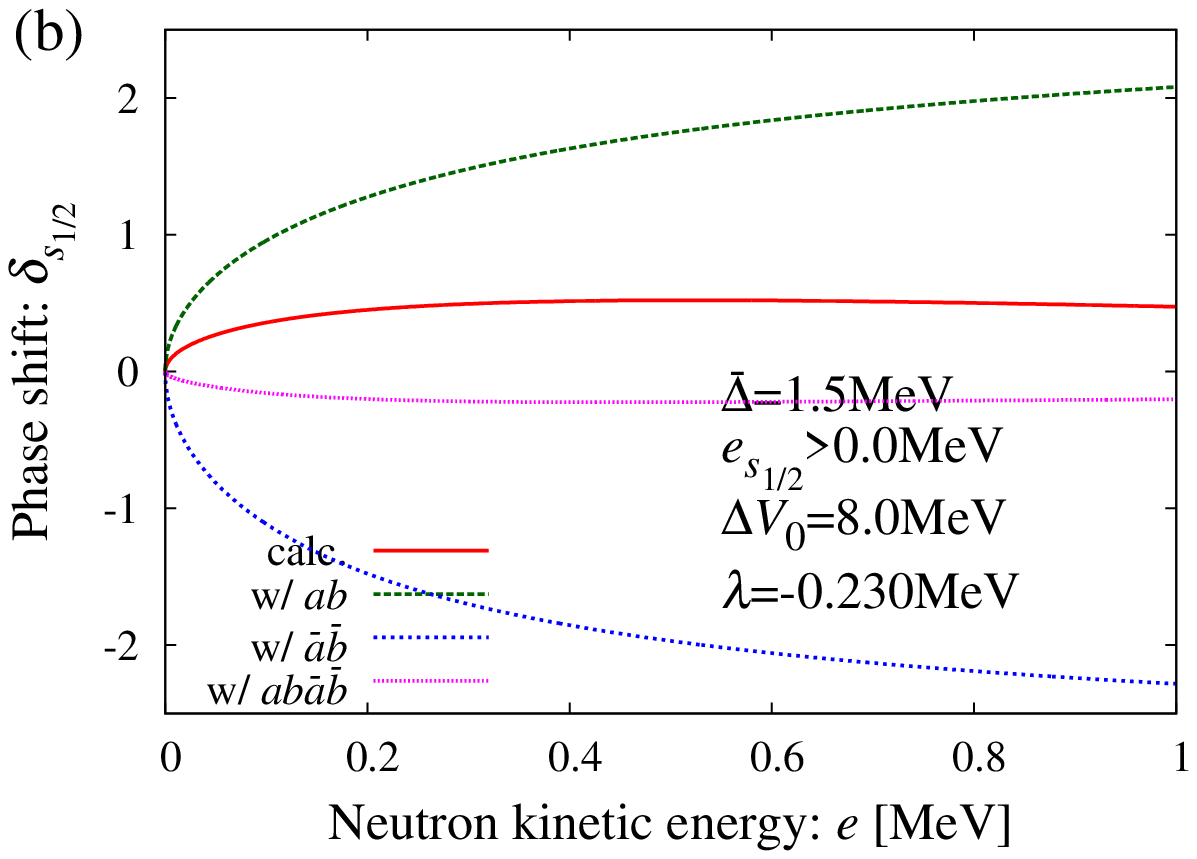}
\end{minipage}
\caption{The same as Fig.~\ref{fig7}, but for $\bar{\Delta}=1.5$ MeV and $\Delta V_0=8.0$ MeV, with which the $2s_{1/2}$ single-particle state is unbound (being a virtual state).}
\label{fig15}
\end{figure}

Figure~\ref{fig14} shows the elastic cross section and the phase shift for various values of $\bar{\Delta}=0.0-3.0$ MeV. The divergent behavior in the cross section around $e\approx 0$, most significant for $\bar{\Delta}=0.0$, tends to diminish as $\bar{\Delta}$ increases. This is because the distance of the pole $a$ from the threshold becomes progressively larger with increasing $\Delta V_0$ ($e_{2s_{1/2}}$). Note also that not only the pole pair $a$ and $b$ but also the other pair $\bar{a}$ and $\bar{b}$ have sizable contributions (Fig.~\ref{fig15}). This is the same feature observed in the cases 1) and 2).

\subsection{Slightly deeper Fermi energy}
Finally we briefly mention the situation where the target nucleus is slightly off the neutron drip-line, i.e. the case where the Fermi energy $\lambda$ is deeper than the above cases, say $\lambda\approx - (1-2) $ MeV.  This corresponds to a target nucleus which has the one-neutron separation energy of $2\sim 3$ MeV. 

\begin{figure}[h]
\begin{center}
\includegraphics[width=80mm]{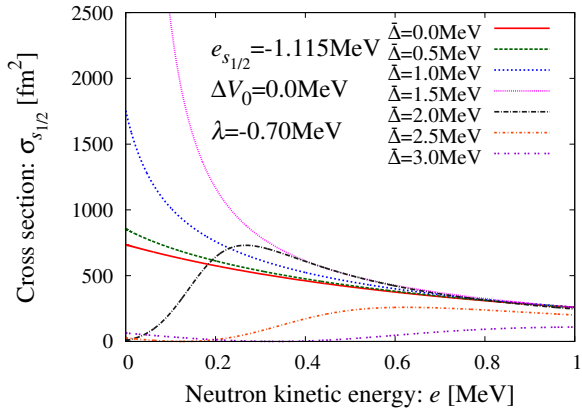}
\end{center}
\caption{Elastic cross section $\sigma_{s_{1/2}}$ in the partial wave $s_{1/2}$ for various values of the average pairing gap $\bar{\Delta}$ with slightly deeper Fermi energy $\lambda=-0.70$ MeV and the neutron $2s_{1/2}$ orbit at $e_{2s_{1/2}}=-1.115$ MeV (with the potential shift $\Delta V_0=0.0$ MeV).}
\label{fig16}
\end{figure}

Let us first discuss the case 2) where the single-particle energy satisfies $2\lambda < e_{2s_{1/2}} < 0$, but here we set the Fermi energy slightly deeper than the case in the previous section. As an example, we choose $\lambda= -0.70$ MeV and $\Delta V_0=0$ MeV ($e_{2s_{1/2}}=- 1.115$ MeV). The elastic cross section obtained with $\bar{\Delta}=0.0,~0.5,~1.0, \cdots 3.0$ MeV is shown in Fig.~\ref{fig16}. Here we observe the characteristic transition from a weakly bound state $\rightarrow$ a virtual state $\rightarrow$ a resonance $\rightarrow$ a broader resonance, which is similar to that of Fig.~\ref{fig1}, where $\lambda= -0.230$ MeV and $\Delta V_0=4.0$ MeV ($e_{2s_{1/2}}=- 0.242$ MeV). With more detailed study we find that the non-perturbative low-energy quasiparticle $s$-wave resonance (such as those in Figs.~\ref{fig16},  \ref{fig1} and~\ref{fig4}) occurs generally if $2\lambda < e_{s_{1/2}} \lesssim \lambda$ and the Fermi energy $\lambda$ is comparable to or smaller than the average pairing gap.

\begin{figure}[h]
\begin{minipage}{0.48\hsize}
\includegraphics[width=77mm]{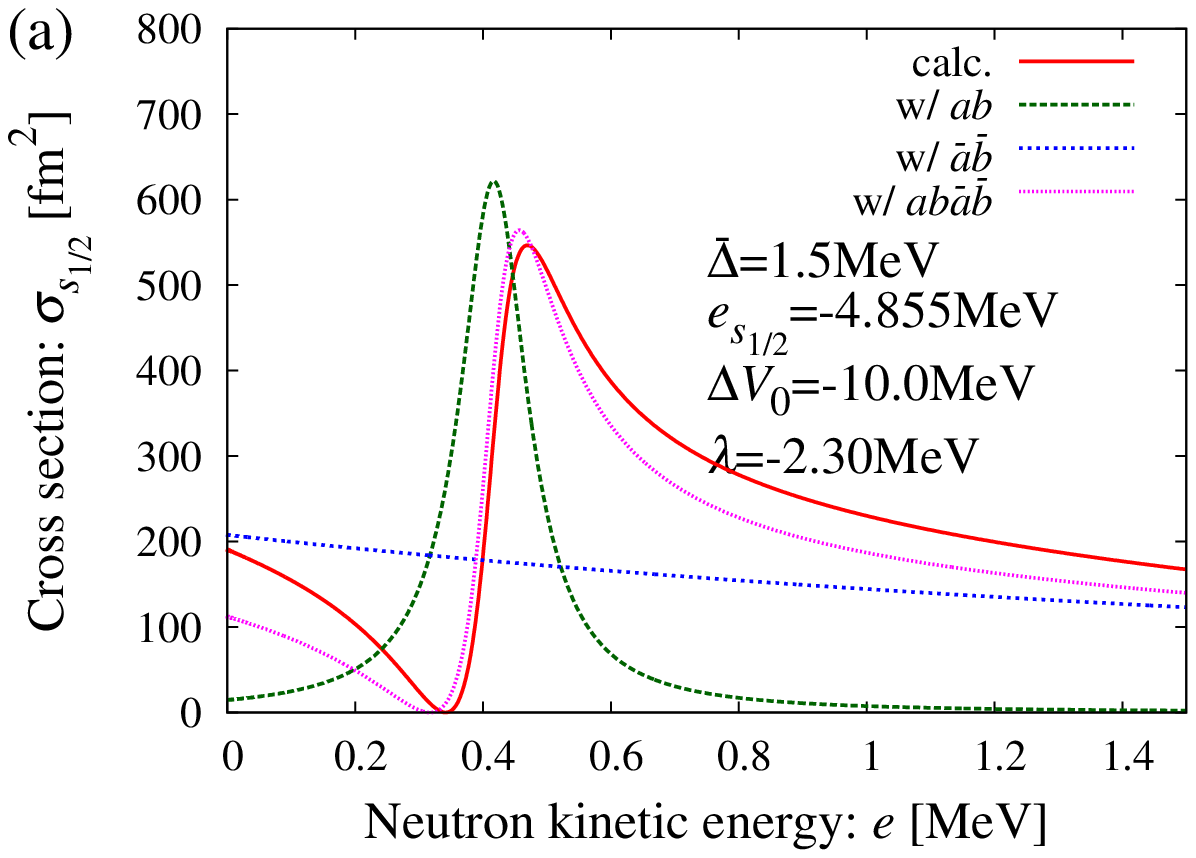}
\end{minipage}
\begin{minipage}{0.48\hsize}
\includegraphics[width=77mm]{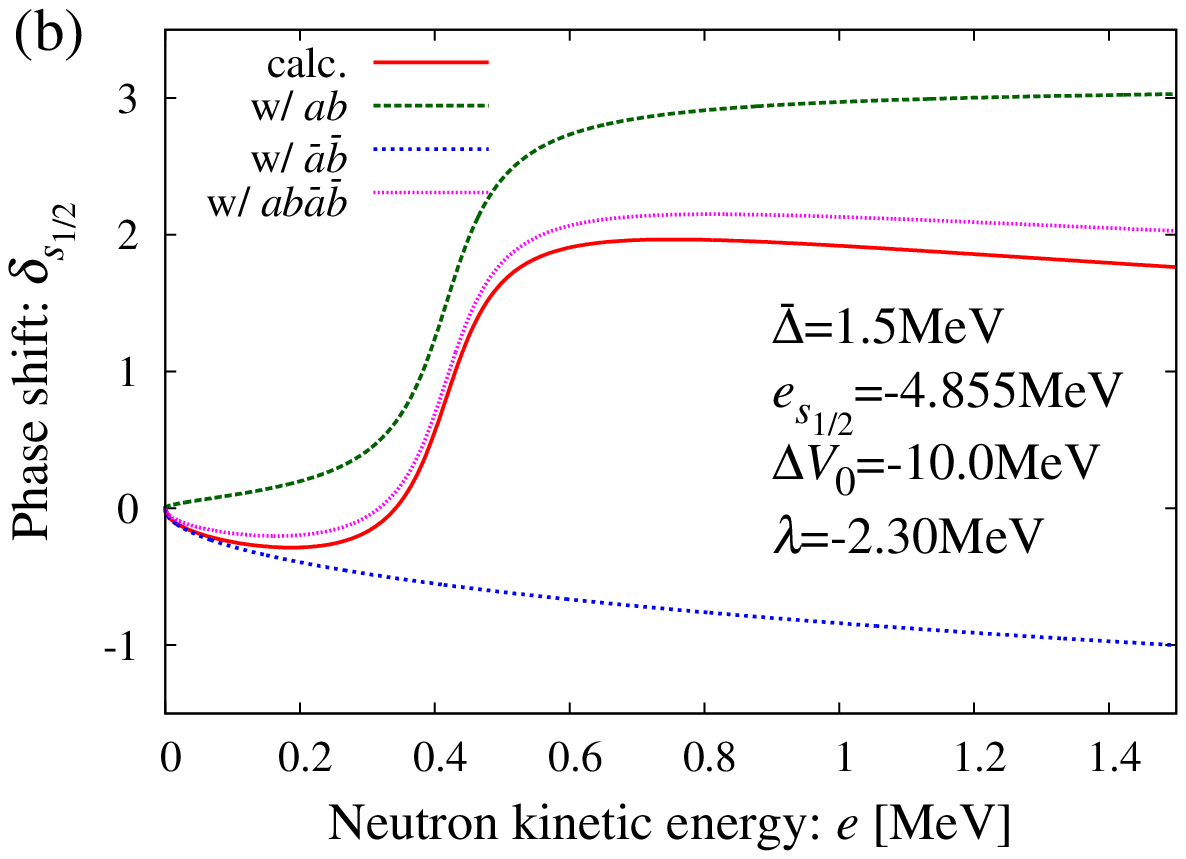}
\end{minipage}
\caption{Influences from S-matrix poles on elastic cross section $\sigma_{s_{1/2}}$ and phase shift $\delta_{s_{1/2}}$ in the case of a moderately deeper Fermi energy $\lambda=-2.30$ MeV with the $2s_{1/2}$ single-particle energy $e_{2s_{1/2}}=-4.855$ MeV ($\Delta V_0=-10.0$ MeV) and the paring gap $\bar{\Delta}=1.5$ MeV.}
\label{fig17}
\end{figure}

Next we consider the case 3) with $e_{2s_{1/2}} < 2\lambda$, where we have the perturbative hole-like quasiparticle resonance (as discussed in Fig.\ref{fig10}), but with a deeper Fermi energy. As an example, we choose a moderately deep Fermi energy $\lambda=-2.30$ MeV, and we rather artificially set $\Delta V_{0}=-10.0$ MeV ($e_{2s_{1/2}}=-4.855$ MeV) to satisfy $e_{2s_{1/2}} < 2\lambda$. The pairing gap is chosen $\bar{\Delta}$= 1.5 MeV, the same as the one in Fig.~\ref{fig12}. In this case the quasiparticle poles $a$ and $b$ appear at $E_{a,b}=2.717 \pm 0.065 i$ MeV; the resonance energy is $e_{a}= \mathrm{Re}\left(E_{a}\right)-|\lambda|=0.417$ MeV, only slightly above the threshold, with a small resonance width $\Gamma_{a}=2\mathrm{Im}\left(E_a\right)=0.129$ MeV, i.e. it appears as a low-energy narrow resonance. Figure~\ref{fig17} shows the elastic cross section and the phase shift. In addition to those obtained with exact numerical evaluation (the red solid curves), the contributions of the quasiparticle poles $a$ and $b$ and the quasihole poles $\bar{a}$ and $\bar{b}$ to the elastic cross section are also shown in Fig.~\ref{fig17}. Here we see that the quasihole poles give only small background in the phase shift, compared with Fig.~\ref{fig12}. The contribution from the quasihole poles $\bar{a}$ and $\bar{b}$ becomes smaller as the Fermi energy becomes deeper since the quasihole poles are located more distant from the threshold than in the case of Fig.~\ref{fig12}. In stable isotopes with deeper Fermi energy, the quasihole poles $\bar{a}$ and $\bar{b}$ become even less important.

\section{Conclusions}
We have investigated properties of unbound single-particle states in pair-correlated drip-line nuclei by describing a low-energy elastic scattering of a neutron in the $s$-wave within the framework of the Bogoliubov's quasiparticle theory. Using a simple model consisting of the Woods-Saxon mean-field and a parametrized pair potential, we solve the Hartree-Fock-Bogoliubov (Bogoliubov-de Genne) equation in the coordinate space with an appropriate boundary condition for the scattering neutron. We have analyzed the elastic cross section and the associated phase shift in terms of the S-matrix poles in order to reveal  a possible resonance structure in the $s$-wave scattering. We focused on  the $2s_{1/2}$ single-neutron orbit which may play a key role in the scattering of $n+ {}^{20,22}$C and a possible $s$-wave quasiparticle resonance in the unbound nucleus ${}^{21,23}$C. In order to explore characteristic features of the $s$-wave neutron scattering and the quasiparticle resonance, we studied systematically by varying the neutron pairing gap and the single-particle energy of the $2s_{1/2}$ orbit.

The S-matrix of the quasiparticle scattering has a four-sheeted Riemann surface as a complex function of the quasiparticle energy. A novel feature is that the S-matrix has four poles which are interpreted as two pairs of poles: one related to a positive-energy quasiparticle state (the quasiparticle poles) and the other originating from a negative-energy quasihole state (the quasihole poles). The quasiparticle poles appear at positions close to the threshold or to the physical region, and hence they influence strongly the low-energy $s$-wave scattering by playing a role of a weakly bound state, a virtual state, or a resonance depending on the single-particle energy and the pairing gap. The other pair, the quasihole poles, also give sizable contributions to the cross section and the phase shift in such a way as if they are another weakly bound single-particle orbit. The four poles (the two pairs of poles) are located at close positions near the threshold because of the shallow neutron Fermi energy (the small neutron separation energy), and consequently their contributions interfere. As the above complex features indicate, the effect of the pairing correlation on the pole positions and on the scattering properties is non-perturbative.

An important feature which may be useful for experimental study is that the pairing correlation gives rise to a low-energy $s$-wave resonance with a large variation in the resonance energy and the resonance width as shown in Fig.~\ref{fig9}. The resonance width varies from zero to $\sim 1$ MeV, and it can be either narrow or wide compared with the resonance energy, depending on the pairing gap and the single-particle energy of the $s$-orbit. A narrow $s$-wave resonance emerges if the $s$-orbit is bound with the single-particle energy well below the Fermi energy. This is a hole-like quasiparticle resonance for which the influence of the pairing correlation is known to be perturbative. A low-energy resonance with moderate width is found to emerge if the $s$-orbit is weakly bound and the Fermi energy is comparable to the pairing gap. In this case the pairing correlation modifies this orbit in the non-perturbative way to produce both a virtual state and a resonance. A resonance pole can be produced also in the case where the mean-field potential has a $s$-wave virtual state, but it may be hard to observe this pole as a resonance since the resonance pole is located far off the physical region in this case.

The present analysis suggests possible presence of a low-energy resonance in the $s$-wave elastic scattering of a neutron on the neutron-rich nucleus ${}^{20,22}$C, or on other unstable nuclei close to neutron-drip line with a neutron $s$-orbit existing around the zero energy. To give a more precise prediction, however, we need to refine the mean-field potential and the pair potential in more quantitative way, including a self-consistent description based on the density functional. It is interesting to study whether the low-energy $s$-wave resonance discussed in the present study can be seen in other observables, e.g. in an invariant mass spectrum of $^{21}$C decaying to the $n+{}^{20}$C channel, populated in nucleon-removal reactions~\cite{Mosby2013,Orr2016,Nakamura2016}. For this purpose, we need to formulate a theoretical framework for the relevant reaction using the Hartree-Fock-Bogolibuov theory. It is also interesting to study the pole structure of the quasiparticle resonance in other theoretical approaches, or to analyze influence of the particle-number violation inherent to the HFB theory. These are subjects of future studies.

\section*{Acknowledgment}

We thank Kenichi Yoshida and Tetsuo Hyodo for fruitful discussions. This work is partially supported by Grant-in-Aid for Research Fellowships of Japan Society for the Promotion of Science (JSPS) for Young Scientists. It is supported also by the JSPS KAKENHI, Grant No. JP17K05436.

\end{document}